\documentclass[aps,prb,preprint,endfloats]{revtex4}
\makeatletter
\def\@dotsep{4.5}
\makeatother
\usepackage{graphicx}
\usepackage{dcolumn}
\usepackage{bm}

\begin{document}


\title{Infrared Surface-Plasmon-Resonance -- a novel biophysical tool for studying living cells}

\author{M. Golosovsky,\footnote{electronic address: golos@vms.huji.ac.il} V. Lirtsman, V. Yashunsky, and D. Davidov}
\affiliation{The Racah Institute of Physics, The Hebrew University of Jerusalem, 91904 Jerusalem, Israel\\}
\author{B. Aroeti}
\affiliation{ Department of Cell and Animal Biology, The Alexander Silberman Institute of Life Sciences,The Hebrew University of Jerusalem, 91904 Jerusalem, Israel\\}
\date{\today}

\begin{abstract} 
We discuss the Surface-Plasmon-Resonance (SPR) technique based on Fourier -Transform - InfraRed (FTIR) spectrometry. We explore the potential of the infrared surface plasmon resonance technique  for biological studies in aqueous solutions and compare it to the conventional surface plasmon technique operating in the visible range.  We demonstrate that the sensitivity of the SPR technique in the infrared range is not lower and in fact is even higher. We show several examples of applying FTIR-SPR for biological studies: (i) monitoring D-glucose concentration  in solution, and (ii) measuring D-glucose uptake by erythrocytes in suspension.  We emphasize the advantages of infrared SPR for studying  living cell cultures and show how this technique can be used for characterization of (i) cholesterol penetration into plasma membrane, and (ii) transferrin-induced clathrin-mediated endocytosis. 
\end{abstract}
\maketitle

\section{General introduction}
Surface plasmon resonance (SPR) is the resonant excitation of the surface electromagnetic wave propagating at the metal-dielectric interface \cite{Raether,Maier2007,Knoll}. This technique  has become an important research tool in biophysics. The interest in it has been spurred by its potential for biosensing, the appearance of  commercial instruments, and the general interest in plasmonics \cite{Maier2005}.  The SPR  technique measures the refractive index or optical absorption with high sensitivity and is particularly advantageous for biosensing  since (i) it is a label-free method \cite{Gauglitz}, and (ii) it can monitor the kinetics of biological processes in real time.  Several overviews summarize  SPR applications in biology and biochemistry  \cite{Homola1999,Homola2003,Homola2008,Phillips,Rich,Haes,Ince,Hoa,Green}. In particular, SPR in the visible range has been used to study membranes  \cite{Alves,Besenicar,Dahlin,Pattnaik,Salamon1997,Salamon2003,Jonsson,Morigaki} and cell cultures \cite{Fang,Giebel,Hide,Yanase}. The SPR technique has been used in several modalities such as SPR-enhanced fluorescence \cite{Liebermann}, SPR imaging \cite{Knoll,Brockman}, and SPR spectroscopy \cite{Kolomenskii,Zangeneh,Zhi,Coe}.

The majority of the SPR applications use laser sources operating in the visible range. Recently, however, several  SPR techniques operating in the near-infrared range have appeared \cite{Brink,Ikehata2003,Ikehata2004,Masson,Patskovsky2003,Patskovsky2004}, some of which use an FTIR spectrometer as a light source \cite{Frutos,Nelson,Lirtsman2005,Ziblat}. In contrast to the conventional SPR technique that operates with a laser source and employs angular interrogation, the FTIR-SPR technique operates with a continuum wavelength source and employs wavelength interrogation.  So far, the FTIR-SPR techniques have utilized glass-based optics that limited its operation to the NIR range. We extended the FTIR-SPR technique to mid-IR range  using a ZnS prism \cite{Lirtsman2008,Yashunsky}. Next,  we will examine new opportunities offered by this technique.

\begin{itemize}
\item \emph{High sensitivity}. Since conductive losses in the infrared range are lower than those in the visible range, the infrared SPR   can be more sensitive than its visible range counterpart.
\item \emph{Penetration depth} of the surface plasmon in the visible range is very short. This is beneficial for studying very thin layers but detrimental for studying cells and cell cultures. The infrared surface plasmon penetrates much deeper and is  more appropriate for studying cells.
\item \emph{Fast multi-wavelength measurements}. The ability to detect SPR at varying wavelengths and/or varying angles allows "tuning" the surface plasmon resonance to any desired spectral range in order to achieve high sensitivity and appropriate penetration depth. 
\item \emph{Spectroscopy}.   Since many biomolecules have specific absorption bands in the infrared range ("fingerprints"), performing multiwavelength SPR measurements in the spectral range of fingerprints, in principle, allows these biomolecules to be identified selectively. 
\end{itemize}
 Next, we will analyze the SPR in the infrared  and identify those niches  where it is especially useful for biosensing. Then we will discuss several  FTIR-SPR applications developed in our group.


\section{Theoretical introduction}

\subsection{Surface plasmon -- definition}
\begin{figure}[ht]
\includegraphics[width=1.5\textwidth]{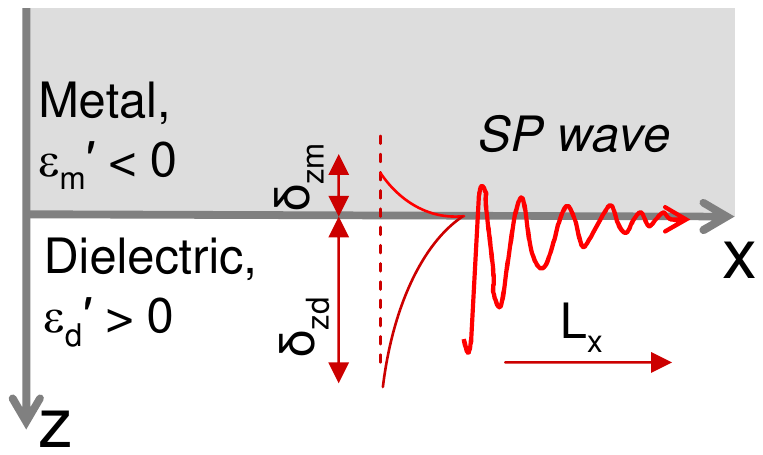}
\caption{Schematic representation of the surface plasmon (SP) propagating at the metal-dielectric interface. $\delta_{zm}$ and $\delta_{zd}$ are  penetration depths into metal and dielectric media, correspondingly; $L_{x}$ is the lateral propagation length that is determined by  conductive, dielectric, and radiation losses.}
\label{fig:SP}
\end{figure}

Surface plasmon (SP) is an electromagnetic wave $\textbf{E}_{sp}=\textbf{E}_{0}e^{-ik_{x}x-k_{z}z}$ that propagates along the metal-dielectric interface and exponentially decays in the perpendicular direction \cite{Raether,Maier2007,Knoll}. The surface plasmon dispersion relations in  the dielectric and in the metal are correspondingly
\begin{eqnarray}
k_{x}^{2}-k_{zd}^{2}=\epsilon_{d}k_{0}^{2}\\
k_{x}^{2}-k_{zm}^{2}=\epsilon_{m}k_{0}^{2} 
\label{dispersion} 
\end{eqnarray}  
Here, $k_{0}$ is  the wave vector in free space, and $\epsilon_{m},\epsilon_{d}$ are the complex dielectric permittivities of the metal and of the dielectric, correspondingly, where $\epsilon_{m}'<0$. When the  surface plasmon propagates along the interface between two unbounded media (Fig. \ref{fig:SP}), its wave vector  is:
\begin{equation}
k_{x}'+ik_{x}''=k_{0}\left(\frac{\epsilon_{m} \epsilon_{d}}{\epsilon_{m}+\epsilon_{d}}\right)^{1/2}
\label{k-x}
\end{equation}
Surface plasmon is a $TM$ wave. The ratio of the longitudinal and the transverse field components is $|E_{z}/E_{x}|=\sqrt{\epsilon_{d}/|\epsilon_{m}|}$ 
in the dielectric medium,  and $|E_{z}/E_{x}|=\sqrt{|\epsilon_{m}|/\epsilon_{d}}$ in the metal.

\subsection{Surface plasmon excitation -- Kretschmann geometry}
\begin{figure}[ht]
\includegraphics[width=1.5\textwidth]{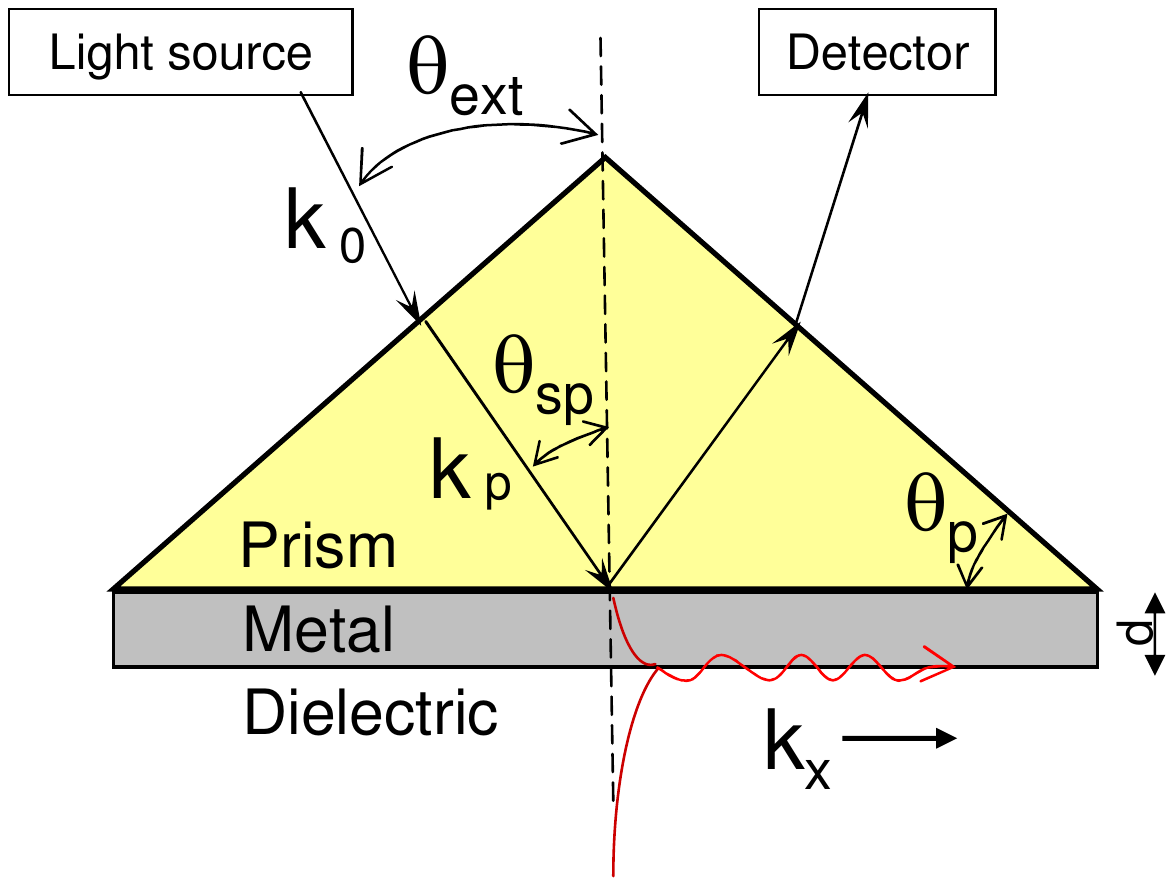}
\caption{Kretschmann geometry of the surface plasmon excitation in the attenuated total reflection regime using a high-refractive-index prism and a thin metal film. $k_{0},k_{p}$ are wave vectors in free space and in the prism; $k_{x}$ is the surface plasmon wave vector. $\Theta_{sp}$ is the angle of incidence at the prism/metal/dielectric interface that corresponds to the resonant excitation of the surface plasmon; $\Theta_{p}$ is the angle at the base of the prism; $\Theta_{ext}$ is the external angle that is determined by $\Theta_{sp}$ and by the prism material and shape. Upon proper choice of the metal film thickness $d$, the resonance reflectivity approaches zero and the off-resonance reflectivity approaches unity.}
\label{fig:prism}
\end{figure}

Because of the limited penetration of the surface plasmon wave into a dielectric medium, it is widely used in monitoring the refractive index of thin dielectric layers.  The most widely used Kretschmann excitation geometry includes  a high-refractive-index prism, a thin metal film,  and a dielectric medium in contact with the this film (Fig. \ref{fig:prism}). The reflectivity as a function of the incident angle (or wavelength) exhibits a sharp minimum at a certain angle $\Theta_{sp}$ corresponding to the surface plasmon excitation \cite{Homola1999}. By measuring the resonant angle/wavelength, one can determine the refractive index of the dielectric medium with high accuracy. The sensitivity of this technique is limited by the angular width of the surface plasmon resonance, $\Delta\Theta_{sp}$. Next, we will develop compact expressions for the SPR angle and width, $\Theta_{sp},\Delta\Theta_{sp}$

\begin{figure}[ht]
\includegraphics[width=0.6\textwidth]{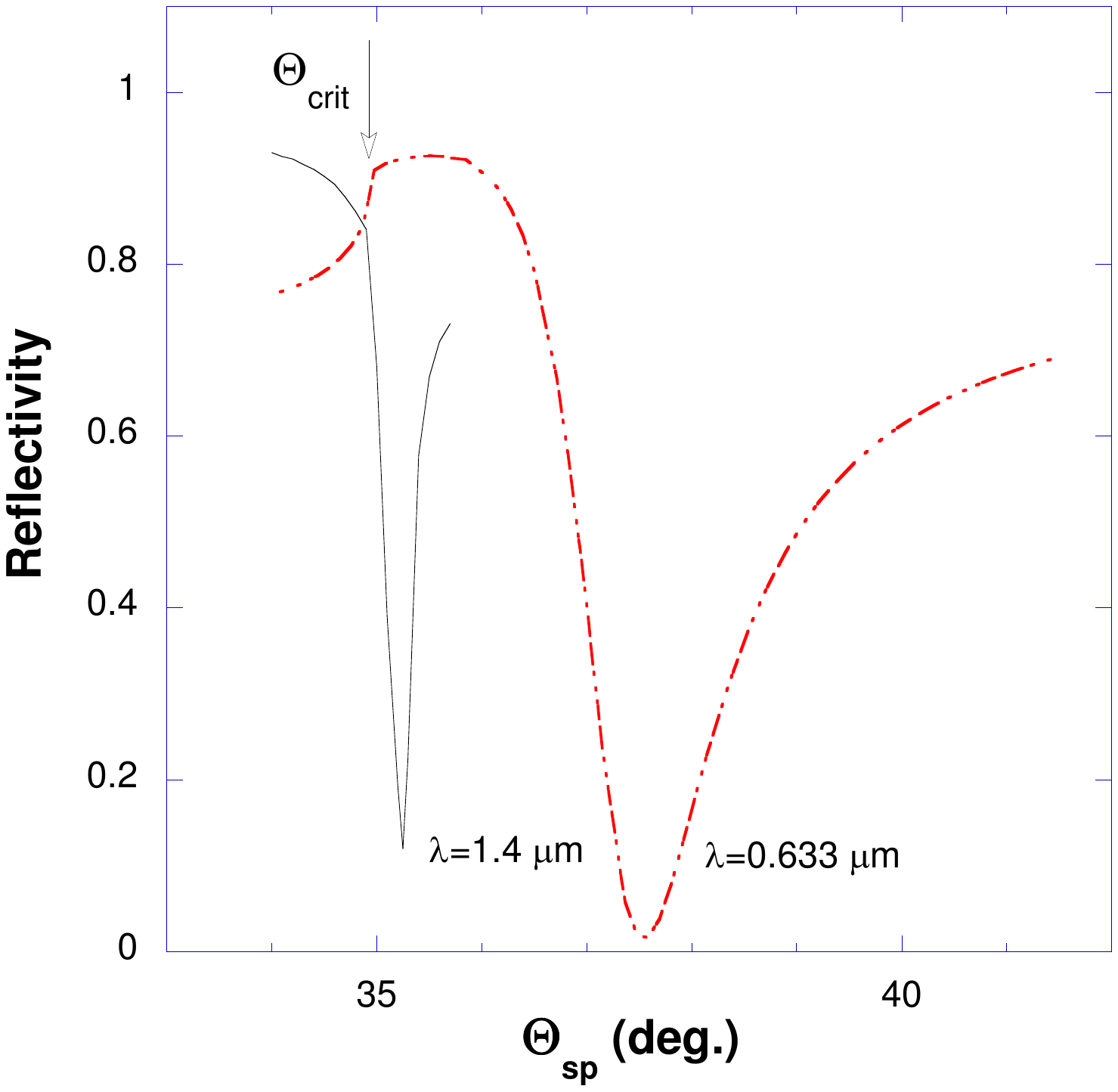}
\caption{Surface plasmon resonance at the Au/air interface. The SPR in the visible range was excited using He-Ne laser and a $60^0$ SF-10  prism coated with a 47-nm-thick Au film.  The SPR in the infrared range was excited using an  FTIR spectrometer and an SF-11 right-angle prism coated with a 30.4-nm-thick Au film. Note the  wide surface plasmon resonance in the visible range ($\lambda=0.633$ $\mu$m) and the sharp resonance in the infrared ($\lambda=1.4$ $\mu$m). The SPR in the visible range is well displaced with respect to the critical angle whereas the SPR in the infrared range is very close to the critical angle.}
\label{fig:example}
\end{figure}

The resonance condition is 
\begin{equation}
k_{x}=k_{p}\sin{\Theta_{sp}},
\label{excitation}
\end{equation}
where $n_{p}$ is  the refractive index and $k_{p}=n_{p}k_{0}$ is the wave vector in the prism.  Equations \ref{k-x},\ref{excitation} yield the resonant angle 
\begin{equation}
\sin{\Theta_{sp}}=\frac{1}{n_{p}}\left(\frac{\epsilon_{m}\epsilon_{d}}{\epsilon_{m}+\epsilon_{d}}\right)^{1/2}.
\label{Theta-SP}
\end{equation}
This angle always exceeds the critical angle
\begin{equation}
\sin{\Theta_{cr}}=\frac{\epsilon_{d}^{1/2}}{n_{p}}.
\label{Theta-cr}
\end{equation}
To recast these expressions using an external angle (see Fig. \ref{fig:prism}), we apply Snell's law and obtain: 
\begin{equation}
\sin{(\Theta_{ext}-\Theta_{p})}=n_{p}\sin{(\Theta_{sp}-\Theta_{p})},
\label{external}
\end{equation}
where s $\Theta_{p}$ is the angle at the base of the prism (Fig. \ref{fig:prism}). Figure \ref{fig:example} compares the SPR in the visible ($\lambda=$0.633 $\mu$m) and in the infrared ($\lambda=$1.4 $\mu$m) range. The SPR appears as a pronounced reflectivity dip. In the visible range, this dip is wide and appears  well above the critical angle, whereas in the infrared range, this dip is very narrow and is very close to the critical angle.

The depth of the surface plasmon resonance in Kretschmann geometry is fine-tuned by the careful choice of the metal film thickness, $d$.  Note that the surface plasmon propagating along a thin metal film is not a true surface wave but rather a leaky wave that radiates a considerable part of its energy  back into the prism \cite{Raether,Johansen}. For the optimal metal film thickness,  the reflectivity at the resonant angle achieves its minimum value and approaches zero (the so-called critical coupling regime), whereas radiation losses are equal to intrinsic losses ($\epsilon_{m}'',\epsilon_{d}''$). The radiation  affects the surface plasmon dispersion relation (Eq. \ref{k-x}) as well. The resonant angle (Eq. \ref{Theta-SP}) is slightly shifted by 1-1.5$^o$, whereas  $k_{x}''$ is changed more significantly: in the critical coupling regime it increases by a factor of two with respect to that given by Eq. \ref{k-x} \cite{Raether,Johansen}.

\subsection{IR surface plasmon resonance propagating along a thin metal film - modelling}
Since we focus on the biological applications of the surface plasmon resonance, we will next consider only water-based samples.   We will assume that the metal film is composed of gold and that its thickness  is optimal, i.e., it corresponds to the critical coupling regime. Since the dielectric constant of gold at infrared frequencies is very high, $\epsilon_{m}'>>\epsilon_{d}\sim 1$, losses  are small, $\epsilon_{m}''/\epsilon_{m}'<<1$. If the dielectric losses are  also small, then $\epsilon_{d}''/\epsilon_{d}'<<1$, then $k_{x}'>>k_{x}''$. Under these conditions, we obtain useful expressions for the real and imaginary parts of the surface plasmon wave vector propagating along the gold film with optimal thickness: 
\begin{eqnarray}
\label{k-x'}
k_{x}'\approx k_{0}\left(\frac{\epsilon'_{m}\epsilon_{d}'} {\epsilon'_{m}+\epsilon_{d}'}\right)^{1/2}\\
\label{k-x''}
k_{x}''\approx k_{0}\left(\frac{\epsilon'_{m} \epsilon_{d}'}
{\epsilon'_{m}+\epsilon_{d}'}\right)^{3/2}
\left(\frac{\epsilon_{d}''}{\epsilon_{d}'^2}+\frac{\epsilon_{m}''}{\epsilon_{m}'^2}\right)\\
\label{k-zm}
k_{zm}=k_{0}\left(\frac{\epsilon_{m}^2}{-\epsilon_{m}-\epsilon_{d}}\right)^{1/2}\\
\label{k-zd}
k_{zd}=k_{0}\left(\frac{\epsilon_{d}^2} {-\epsilon_{m}-\epsilon_{d}}\right)^{1/2}
\end{eqnarray}
(Note that $\epsilon_{m}'<0$.) The surface plasmon penetration depth into the dielectric is
\begin{equation}
\delta_{zd}=\frac{1}{2k_{zd}}=\frac{\lambda_{0}}{4\pi}\left(\frac{-\epsilon'_{m}- \epsilon_{d}'}
{\epsilon_{d}'^{2}}\right)^{1/2}
\label{delta-z}
\end{equation}
The surface plasmon propagation length along the metal-dielectric interface is \cite{Raether,Johansen}
\begin{equation}
L_{x}=\frac{1}{2k_{x}''}=\frac{\lambda_{0}}{4\pi}\left(\frac{\epsilon'_{m}+ \epsilon_{d}'}
{\epsilon'_{m}\epsilon_{d}'}\right)^{3/2}
\left(\frac{\epsilon_{d}''}{\epsilon_{d}'^2}+\frac{\epsilon_{m}''}{\epsilon_{m}'^2}\right)^{-1}
\label{L-x}
\end{equation}
Since the dielectric constant of the metal  in the infrared range is very high, $|\epsilon_{m}'|>>1$, then $\epsilon_{m}''/\epsilon_{m}'^2<<1$ (this is valid even if the loss tangent of the metal is high, $\epsilon_{m}''/|\epsilon_{m}'|\leq 1$). Therefore, the conductor loss contribution to $L_{x}$ is negligible and it is determined mainly by the dielectric loss. 

To estimate the surface plasmon resonance width, $\Delta\Theta_{sp}$, we noted that Eq. \ref{excitation} yields $\delta k_{x}=\cos{\Theta_{sp}}\delta\Theta_{sp}$. We replaced $\delta k_{x}$ by $k_{x}''$.  Thus,  
\begin{equation}
\Delta\Theta_{sp}=A \frac{k_{x}''}{k_{p}\cos{\Theta_{sp}}},
\label{width}
\end{equation}
where $A$ is a numerical constant that depends on how the SPR width is defined (FWHM or differently). Using Eqs. \ref{external},\ref{width}, we obtain
\begin{equation}
\Delta\Theta_{ext}=A\frac{\cos{(\Theta_{p}-\Theta_{sp})}}
{\cos{(\Theta_{p}-\Theta_{ext})}\cos{\Theta_{sp}}}
\frac{k_{x}''}{k_{0}}
\label{width-ext}
\end{equation}
For realistic parameters of the prism and the sample ($n_{p}=2-2.3, \Theta_{p}=45^0, n_{d}=1.25-1.45$), the angular dependence of the prefactor, $\frac{\cos{(\Theta_{p}-\Theta_{sp})}}
{\cos{(\Theta_{p}-\Theta_{ext})}\cos{\Theta_{sp}}}$, is very weak and it can be replaced by a numerical constant. Using Eq. \ref{k-x''}, we recast Eq. \ref{width-ext} as follows:
\begin{equation}
\Delta\Theta_{ext}\approx \tilde{A}\epsilon_{d}^{3/2}
\left(\frac{\epsilon_{m}''}{\epsilon_{m}'^2}+\frac{\epsilon_{d}''}{\epsilon_{d}'^2}\right),
\label{width-ext1}
\end{equation}
where $\tilde{A}$ absorbs all numerical constants. A comparison of the exact numerical calculations for the ZnS prism and water as a dielectric medium yields $\tilde{A}=1.85$. An approximate expression for the SPR width is:
\begin{equation}
\Delta\Theta=\sqrt{\Delta\Theta_{ext}^2+\Delta\Theta_{div}^2},
\label{divergence}
\end{equation}
where $\Delta\Theta_{div}$ is the incident beam divergence. Although the SPR width can be calculated exactly using Fresnel reflectivity formulae, Eqs. \ref{width-ext1},\ref{divergence} are more amenable to  analysis. 

The SPR width is a crucial factor that determines the sensitivity of the SPR technique. The latter  is  usually \cite{Homola1999,Johansen} defined as  $\frac{\partial R}{\partial n}$, where $R$ is the reflectivity and $n=\epsilon_{d}^{1/2}$.  After recasting this expression using an angular variable, we obtain $S_{bulk}=\frac{\partial R}{\partial \Theta_{ext}}\frac{\partial \Theta_{ext}}{\partial n}$. The maximum sensitivity is achieved at some angle that corresponds to the sloping edge of the surface plasmon resonance, where reflectivity sharply drops from unity to almost zero. Hence 
\begin{equation}
S_{bulk}\approx \frac{1}{\Delta \Theta}\frac{\partial \Theta_{ext}}{\partial n}
\label{S-bulk}
\end{equation}
The first factor here is determined by the SPR width  (Eqs. \ref{width-ext},\ref{divergence}), whereas the second factor, $\frac{\partial \Theta_{ext}}{\partial n}$, is a smooth function of the refractive index that is derived from Eqs. \ref{Theta-SP},\ref{external}. 

To characterize  thin dielectric layers with a thickness smaller than the surface plasmon penetration depth $\delta_{zd}$  (Eq. \ref{delta-z}, the SPR sensitivity is defined  as follows \cite{Johansen},
\begin{equation}
S_{layer}=\frac{S_{bulk}}{\delta_{zd}}.
\label{S-layer}
\end{equation}

\subsection{IR surface plasmon resonance at the ZnS/Au/water interface - practical aspects} 
We will now consider practical aspects of the infrared surface plasmon resonance in the biological, i.e., water-based samples. Although SPR at infrared frequencies can be excited in many metals \cite{Sambles},  the requirement of biocompatibility leaves gold as the only choice.  For the near IR, one can use  BK-7 or SF-11 glass prisms, whereas for  mid-IR, one can use ZnS or ZnSe prisms. 

\begin{figure}[ht]
\includegraphics[width=0.6\textwidth]{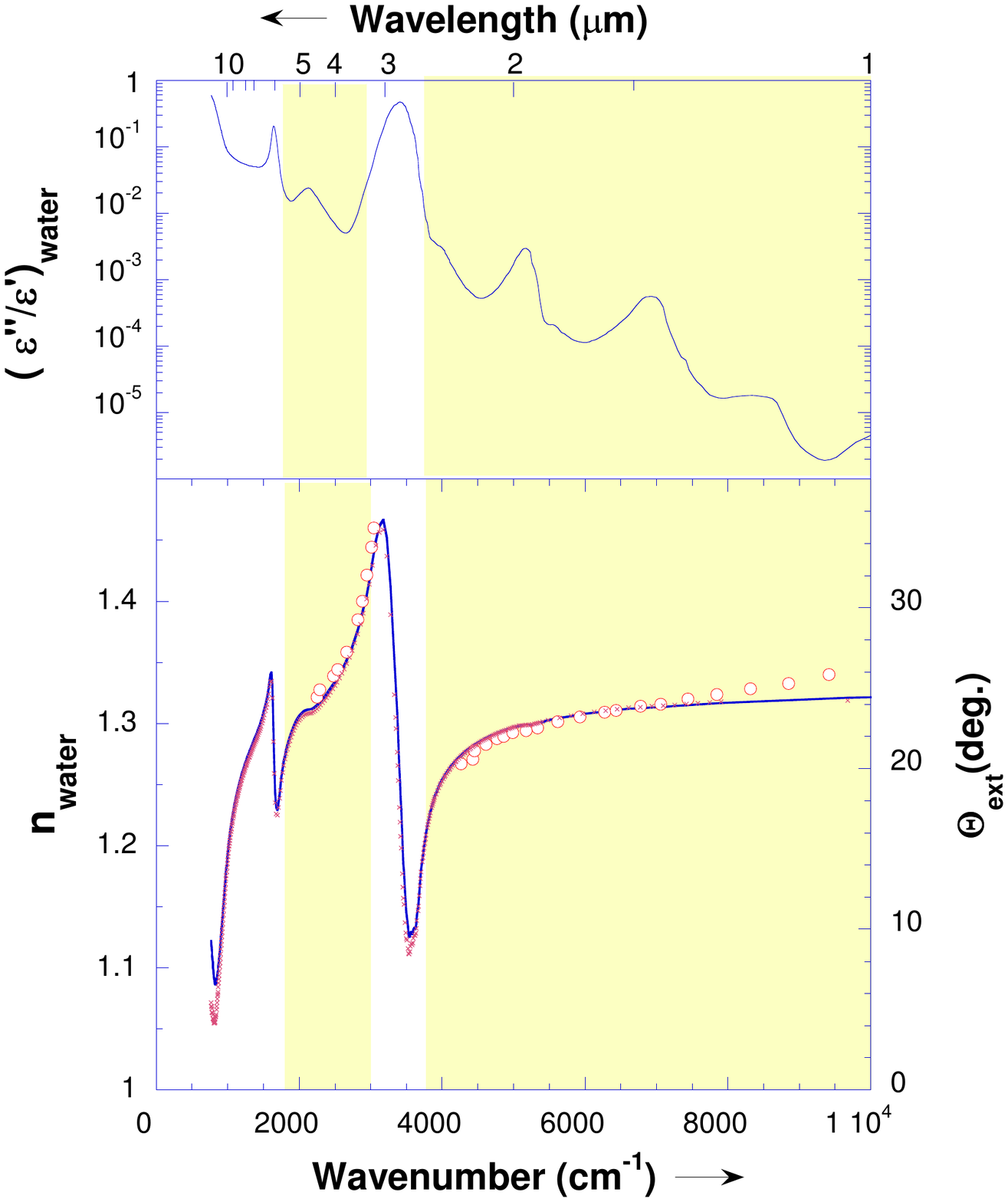}
\caption{Upper panel: loss tangent of water. The colored areas indicate spectral windows where water absorption is sufficiently low and the surface plasmon can be excited. Lower panel: the continuous blue line shows the refractive index of water. Purple crosses denote the SPR angle, $\Theta_{ext}$, for the ZnS/Au/water interface, as given by Eqs. \ref{Theta-SP},\ref{external} (right $y$-axis). Red circles indicate our experimental results for a 18-nm-thick Au film. The wavelength dependence of the SPR angle mimics the water refractive index.}
\label{fig:water}
\end{figure}

The range of wavelengths/angles where  SPR can be excited is determined by the dielectric losses in water and by the conductive losses in the gold film. Figure \ref{fig:water} shows  dielectric losses (upper panel) and  the refractive index  of water (lower panel).  It is clear that SPR at the Au/water interface can be excited only in several "allowed" spectral windows where water absorption is sufficiently low. Figure \ref{fig:water}  shows the SPR angle vs. the wavelength, as predicted by Eqs. \ref{Theta-SP},\ref{external} (lower panel, right $y$-axis). This dependence  mimics the refractive index of water. Note that in the angular interval, $\Theta_{ext}= 20-25^o$, there are two surface plasmon resonances  at each angle.

\begin{figure}[ht]
\includegraphics[width=0.6\textwidth]{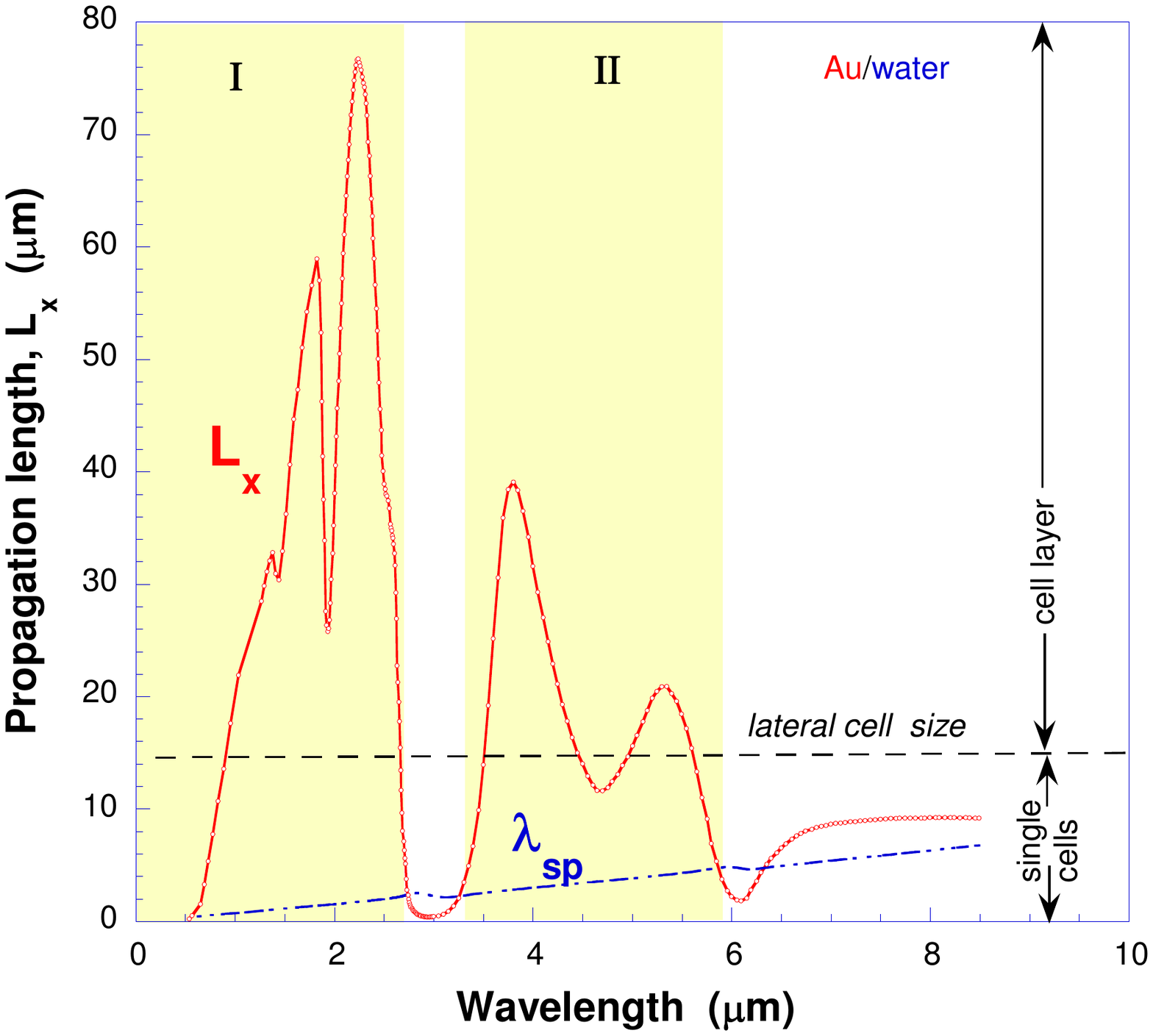}
\caption{The open circles denote surface plasmon propagation length, $L_{x}$, at the ZnS/Au/water interface, as given by Eq. \ref{L-x} for the optimal Au film thickness. The dash-dotted blue line denotes surface plasmon wavelength, $\lambda_{sp}$. The colored areas indicate spectral windows where SPR can be excited (see Fig. \ref{fig:water}). In the "allowed" spectral windows (0.5-3 $\mu$m and  3.3-6 $\mu$m) $L_{x}>\lambda_{sp}$, whereas in the "forbidden" windows (3-3.3 $\mu$m and above 6 $\mu$m), $L_{x}\leq\lambda_{sp}$. The dashed line denotes lateral cell size. When  surface plasmon propagates through the cell monolayer grown on gold, it senses either single cells (spectral windows of  0.5-1 $\mu$m and 4.5-6 $\mu$m -see Fig. \ref{fig:cells}e) or some effective medium consisting of cells and growth solution (spectral windows of  1-2.7 $\mu$m and 3.5-4.5 $\mu$m)}
\label{fig:propagation-length}
\end{figure}

Figure \ref{fig:propagation-length} shows surface plasmon propagation length, $L_{x}$, as predicted by Eq. \ref{L-x}. This length should be compared to the surface plasmon wavelength, $\lambda_{sp}=2\pi/k_{x}'$. In the "allowed" spectral ranges, the surface plasmon attenuation is low and surface plasmon propagates to long distances, $L_{x}/\lambda_{sp}>>1$. In the "forbidden" spectral ranges, the surface plasmon resonance is overdamped and the surface plasmon wave quickly attenuates, yielding $L_{x}/\lambda_{sp}<~1$.    

In the context of sensing biological cell cultures, it is useful to compare the surface plasmon propagation length, $L_{x}$, to the  lateral size of the cell, $d_{cell}\sim 10-20$ $\mu$m.  In the spectral ranges  $\lambda <1.2$ $\mu$m and   $\lambda >4.5 \mu$m, the surface plasmon propagation length is small, $L_{x}<d_{cell}$, and reflectivity in the SPR regime is the sum of the contributions of individual cells. In the spectral ranges $\lambda_{sp}=1.2 -2.6$ $\mu$m and $\lambda_{sp}\approx 4$ $\mu$m, the surface plasmon propagation length is long, $L_{x}>d_{cell}$; hence, the surface plasmon wave senses an "effective medium"  consisting of cells and gaps between the cells.

\begin{figure}[ht]
\includegraphics[width=0.6\textwidth]{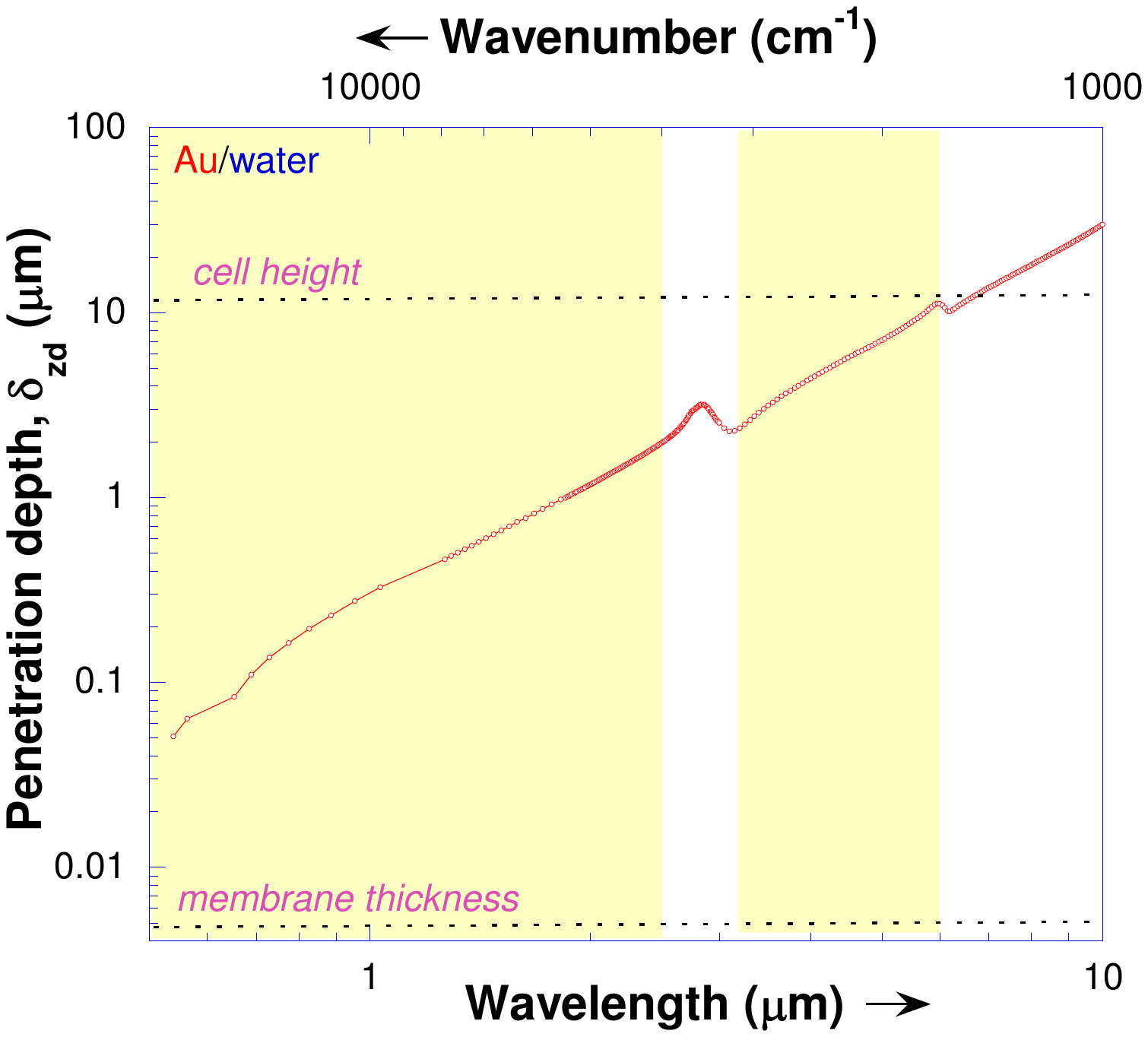}
\caption{Surface plasmon penetration depth into biolayer (aqueous solution) as defined by Eq. \ref{delta-z}. The colored areas show spectral windows where SPR at the Au-water interface can be excited. In studying cell monolayers, it is important to compare penetration depth (open circles), cell height, and membrane thickness (dashed lines). Penetration of the short-wavelength  surface plasmon by short wavelengths  into the cells is strongly limited; hence, it is mostly sensitive to events related to the cell membrane. The long-wavelength surface plasmon penetrates deeper into cells and senses the whole cell volume.}
\label{fig:penetration-depth}
\end{figure}

Figure \ref{fig:penetration-depth} shows surface plasmon penetration depth into water-based dielectric medium. In the visible and near-infrared ranges, it is exceedingly small, $\delta_{z}< 0.24$ $\mu$m. In the context of cell cultures grown on gold, such a small penetration depth  is advantageous for studying the cell membrane and cell adhesion. However, a small penetration depth precludes studying the cell interior. In the spectral range $\lambda= 2-2.5$ $\mu$m, the surface plasmon penetration depth is $\delta_{zd}=1-2$ $\mu$m. Hence, the surface plasmon wave penetrates deep enough to study events in the cell cytoplasm but not in the nucleus.  For the wavelength range $\lambda=4-6$ $\mu$m, the penetration depth is comparable to the cell size; hence, the whole cell volume is accessible for the surface plasmon. 

\begin{figure}[ht]
\includegraphics[width=0.6\textwidth]{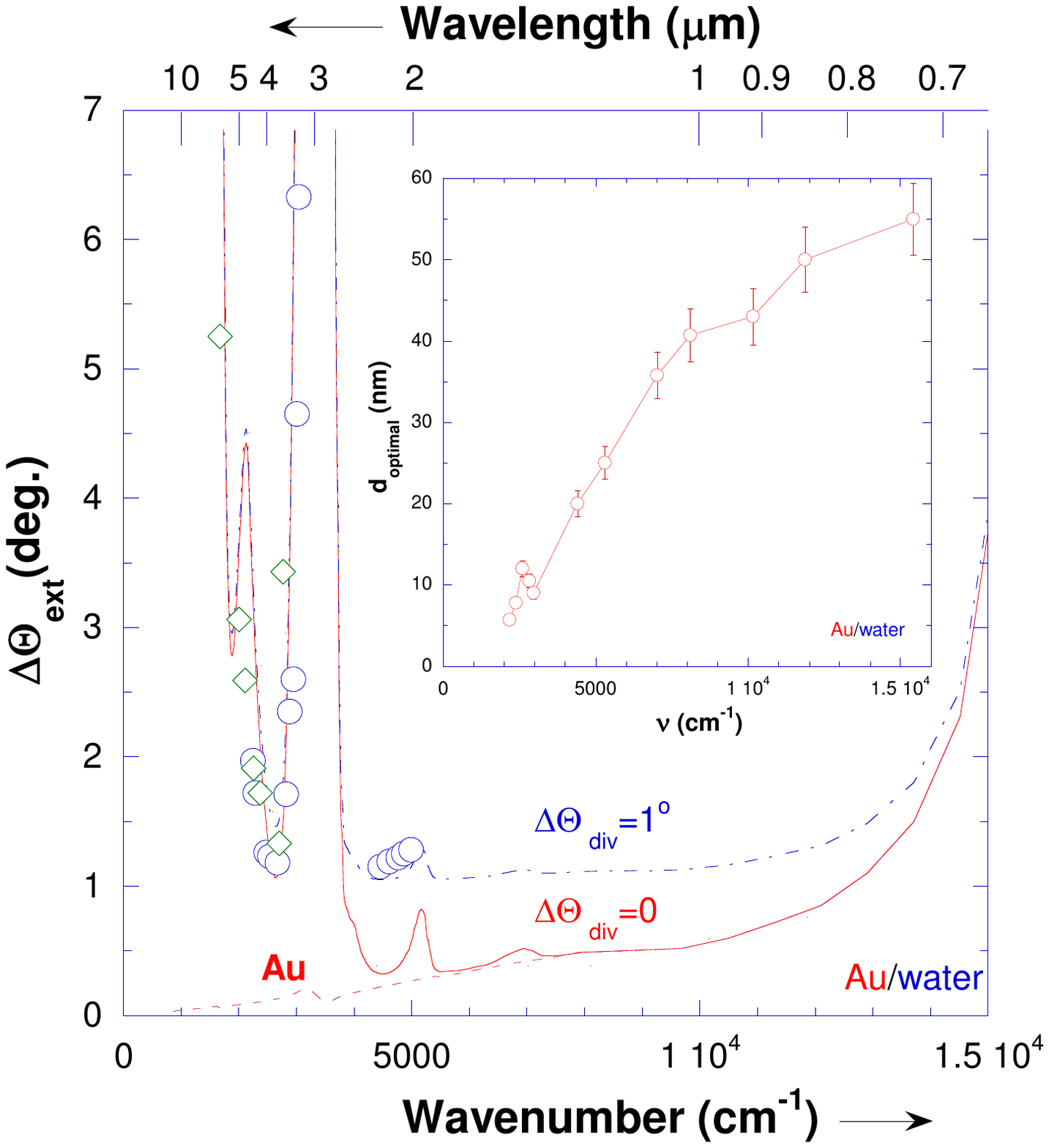}
\caption{Surface plasmon resonance width for the ZnS/Au/water interface.  The lines  show SPR width, as determined by the absorption in water and in Au and by the coupling (radiation) losses, as predicted by Eq. \ref{width-ext1}. The continuous red line denotes zero beam divergence and the dash-dotted blue line denotes a beam divergence of $1^0$. The dashed red line indicates the prediction of Eq. \ref{width}, assuming negligible losses in water.  In all calculations, we assumed that the Au film thickness is optimal, i.e., it satisfies the critical coupling condition at a given wavelength. The inset shows the optimal Au film thickness that was determined from numerical simulations based on Fresnel reflectivity formulae. Blue circles indicate our experimental results obtained with a 14-nm-thick  Au film. The beam divergence was $\Delta\Theta_{div}= 0.8^{0}$.}
\label{fig:width}
\end{figure}

Figure \ref{fig:width} shows the SPR width  vs. wavelength, as given by  Eqs. \ref{width-ext1},\ref{divergence}. The exact calculation, using Fresnel reflectivity formulae,  yields almost the same results.  Note that in the near-infrared range ($\lambda<1.3$ $\mu$m), where water optical absorption is negligible, the SPR width is determined by conductor losses.  On the other hand, at $\lambda> 2.5$ $\mu$m, conductor losses are negligible and the SPR width is determined by water absorption. In the intermediate range, $\lambda=1.3-2.5$ $\mu$m,  the SPR width achieves its minimum value of 0.3$^o$, whereas the contributions of water absorption and conductor losses are comparable. However, to take full advantage of such a narrow resonance, the incident beam should be very well collimated, otherwise the SPR width in this spectral range would be limited by beam divergence.

The inset in  Fig. \ref{fig:width} shows the optimal gold film thickness, $d$, that is required to achieve critical coupling. $d$ decreases with increasing wavelength, in such a way that the operation in mid-IR requires 10-12-nm-thick gold films; this  approaches the practical limit at which continuous gold films can be fabricated. When the film thickness deviates from its optimal value, the SPR is  excited,  although the SPR minimum is not so deep \cite{Raether}. The symbols in Fig. \ref{fig:width} show our experimental results achieved in two experiments with a 12-nm-thick gold film that was optimized for the operation at $\lambda=5$ $\mu$m. At $\lambda>3$ $\mu$m, we observe a good correspondence between the approximate model and experimental SPR values, whereas at $\lambda<2.5$ $\mu$m, the experimental SPR is considerably broadened due to beam divergence and deviation of the film thickness from the optimal value corresponding to this wavelength. 

\begin{figure}[ht]
\includegraphics[width=0.8\textwidth]{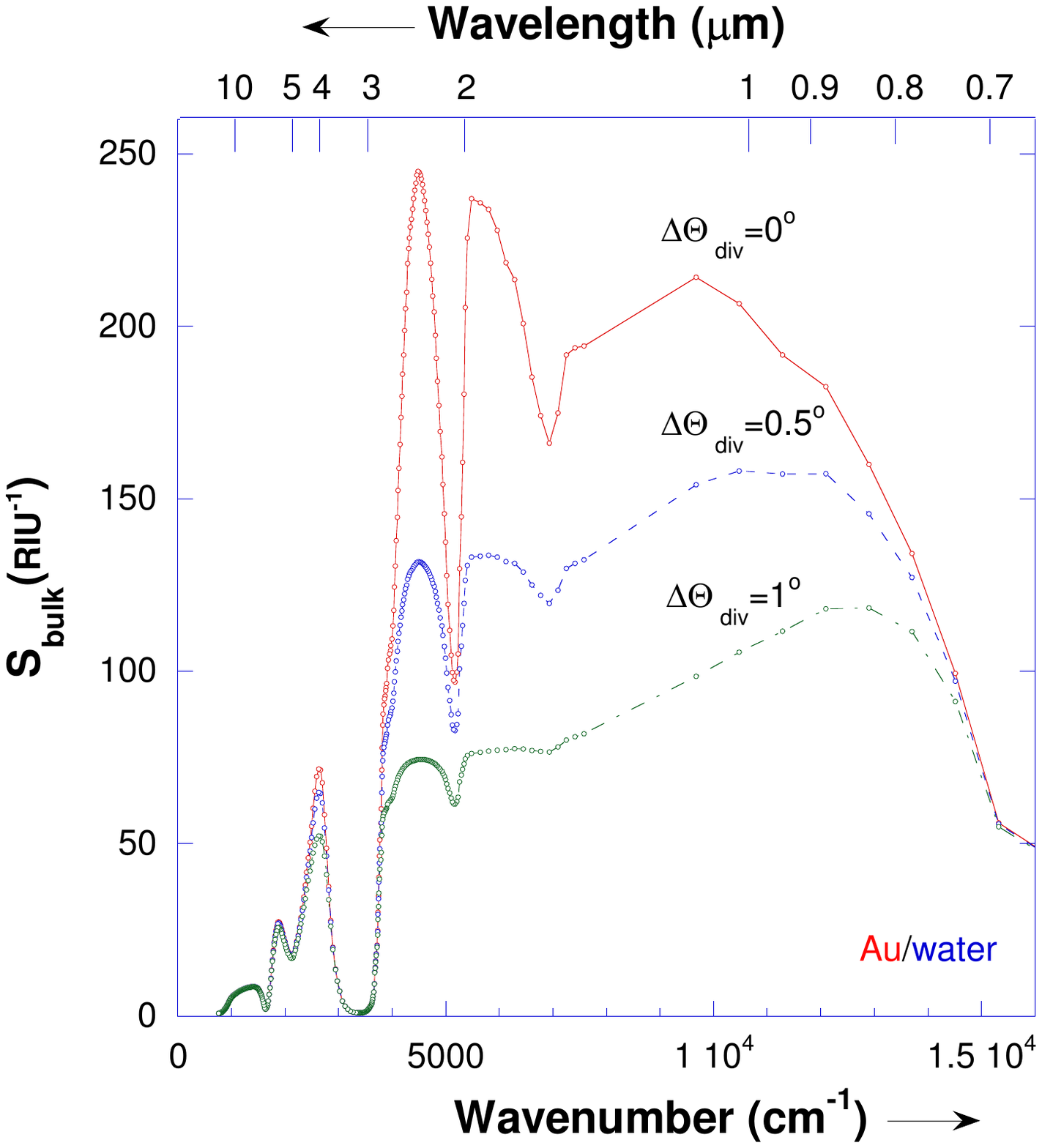}
\caption{The open circles denote the calculated sensitivity of the SPR technique, $S_{bulk}=\partial R/\partial n$, for different beam divergences, as defined by Eq. \ref{S-bulk} and Fig. \ref{fig:water}. The lines guide the eye. (Our calculation does not extend into spectral region between 7670 cm$^{-1}$ and 9670 cm$^{-1}$ owing to uncertainty in the optical parameters of Au arising from the mismatch between two data sets \cite{Palik}.) Sensitivity in mid-infrared is higher than that in the visible range.}
\label{fig:sens-bulk}
\end{figure}

Figure \ref{fig:sens-bulk} shows the bulk sensitivity of the SPR technique estimated according to Eq. \ref{S-bulk}. The bulk sensitivity in the infrared range is considerably higher than that in the visible range. In the range $\lambda=0.6-2.5$ $\mu$m, the sensitivity is more or less constant, whereas in the long-wavelength window, $\lambda=3.3-5$ $\mu$m, it is lower but  is still comparable to that in the visible range. This means that both these spectral windows may be used for SPR-spectroscopy.

\begin{figure}[ht]
\includegraphics[width=0.8\textwidth]{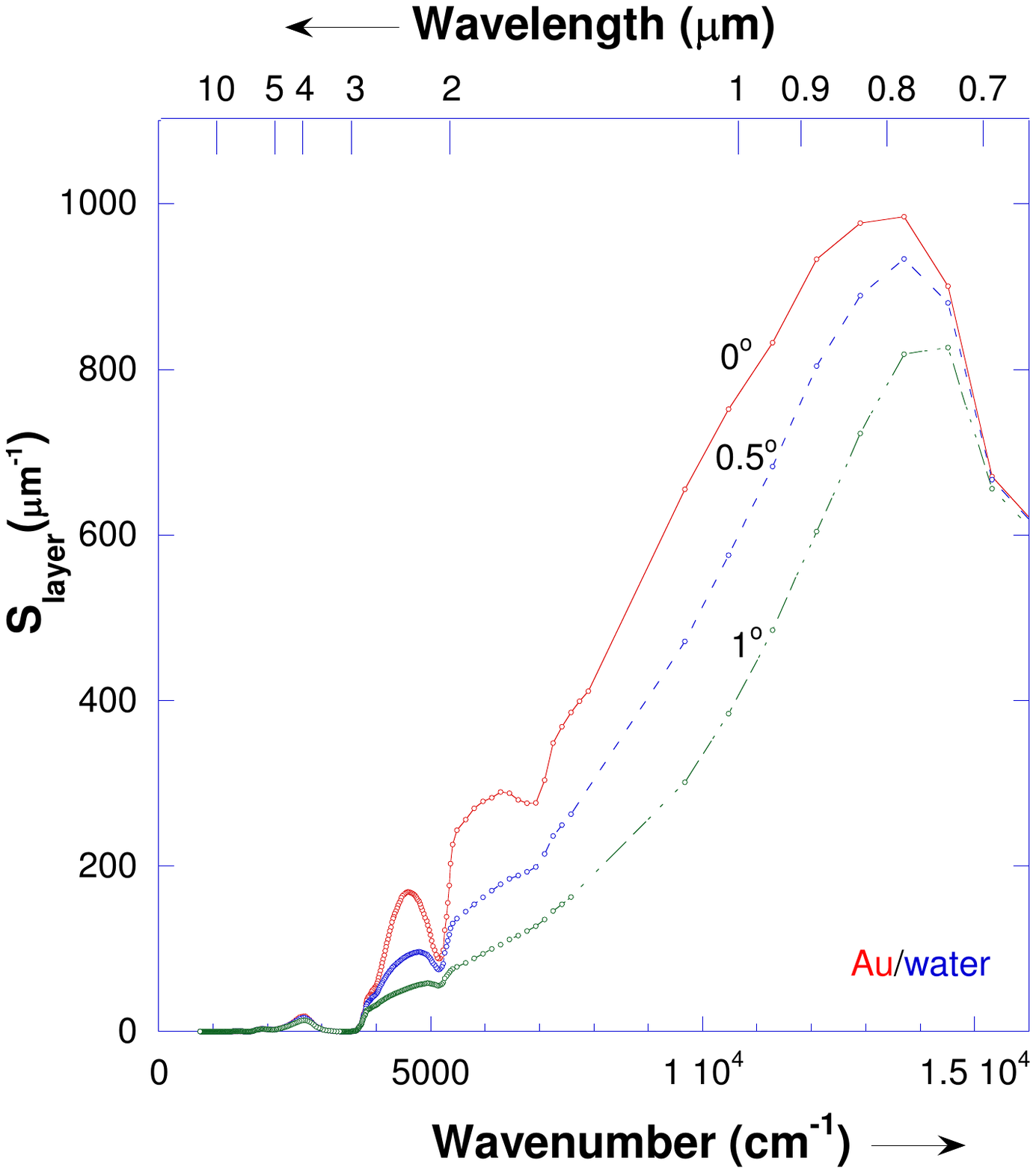}
\caption{Sensitivity of the SPR technique for a thin layer with thickness much smaller than the SP penetration depth,  as defined by Eq. \ref{S-layer} and Fig. \ref{fig:water}. The number at each curve indicates beam divergence. The maximum sensitivity is achieved  for $\lambda\sim 0.8$ $\mu$m.} 
\label{fig:sens-layer}
\end{figure}

Figure \ref{fig:sens-layer} shows the thin layer sensitivity. The maximum sensitivity is achieved in NIR, at $\lambda=$ 800 nm, whereas the sensitivity in the mid-IR is lower. Hence, the NIR range is especially advantageous for studying thin dielectric films on gold substrate. 

\subsection{Surface plasmon propagation in biological samples}
\begin{figure}[ht]
\includegraphics[width=0.8\textwidth]{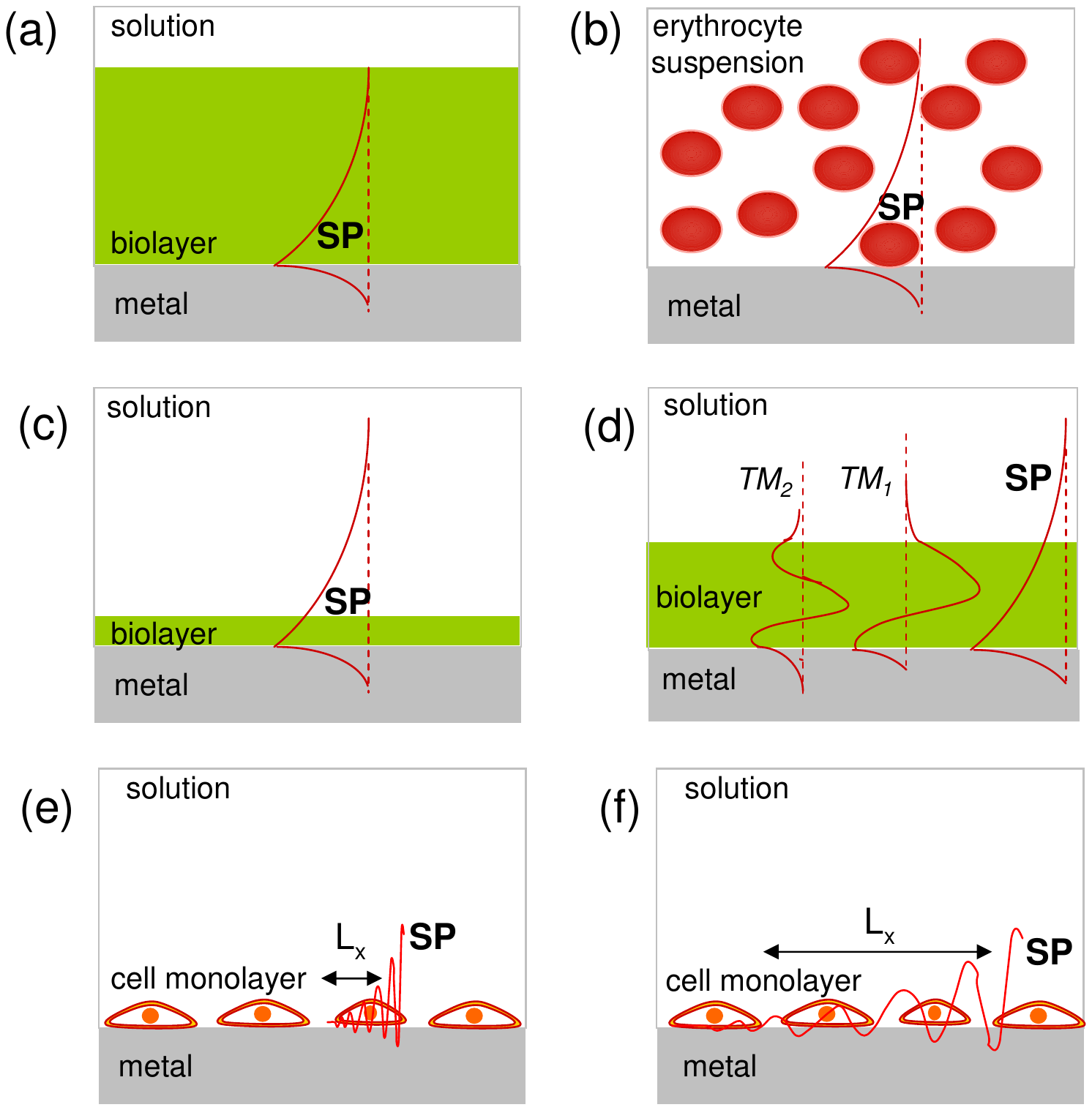}
\caption{Surface plasmon propagation in biological samples. (a) Thick biolayer (solution or bulk sample).  Surface plasmon  decays within this layer and is not sensitive to anything beyond it. (b) Cell suspension, for example, erythrocytes. Surface plasmon partially or fully penetrates into cells that are close to the metal. SP can be used to study these cells provided the SP penetration depth is comparable to the cell size. (c) A very thin biolayer in contact with metal (membrane, adsorbed molecules). To study such thin layers, the SP penetration depth should be short enough. (d) A biolayer of intermediate thickness in contact with metal. Here, in addition to the exponentially decaying SP wave, the guided modes  can be excited as well. These waveguide modes have sinusoidal field distribution inside the layer and may be advantageous for layer characterization. (e) Cell culture grown on metal. The SP propagation length is smaller than the lateral cell size. The reflectivity in the SPR regime represents a sum of the contributions of individual cells. (f) SP propagation length exceeds lateral cell size. The SP probes an "effective medium" consisting of cells and extracellular space.} 
\label{fig:cells}
\end{figure}

We consider several regimes in which the FTIR-SPR technique can be used to study biological objects. These objects (cell membrane, single cell, cell culture, etc.) have different sizes and the surface plasmon propagation there is also different. Indeed, the SP propagation depends on the relation between the spatial extent of the SP wave and the characteristic size of the sample. Hence, by selecting the proper wavelength, the SPR can be made especially sensitive to the processes occurring at a targeted spatial scale. Figure \ref{fig:cells} illustrates several regimes of surface plasmon sensing. 
\begin{itemize}
    \item (a,b) Thick and macroscopically uniform biolayers. Here, SP probes the bulk  refractive index of the biolayer. This can be used for monitoring the biologically important molecules in solutions, for example, glucose. An interesting modification is a suspension of floating cells,  for example, erythrocytes.

\item (c) A very thin biolayer attached to the gold film, for example, a membrane, or biomolecular recognition layer with a thickness of only 1-5 nm. Although the surface plasmon penetration depth usually exceeds this thickness, the short-wavelength surface plasmon is sensitive enough to characterize them.

 \item (d) A moderately  thick biolayer  with thickness  comparable to the wavelength. Here, in addition to the surface plasmon wave, additional waveguide modes can appear. These  can be useful for depth-resolved biosensing,  and for probing  cell adhesion and cell-cell attachment.

 \item (e,f) Cell cultures grown directly on metal.  Here, depending on the propagation length, the surface plasmon wave can probe either individual cells or the cell layer as a whole.
\end{itemize}


\section{Experimental}
\begin{figure}[ht]
\includegraphics[width=1.2\textwidth]{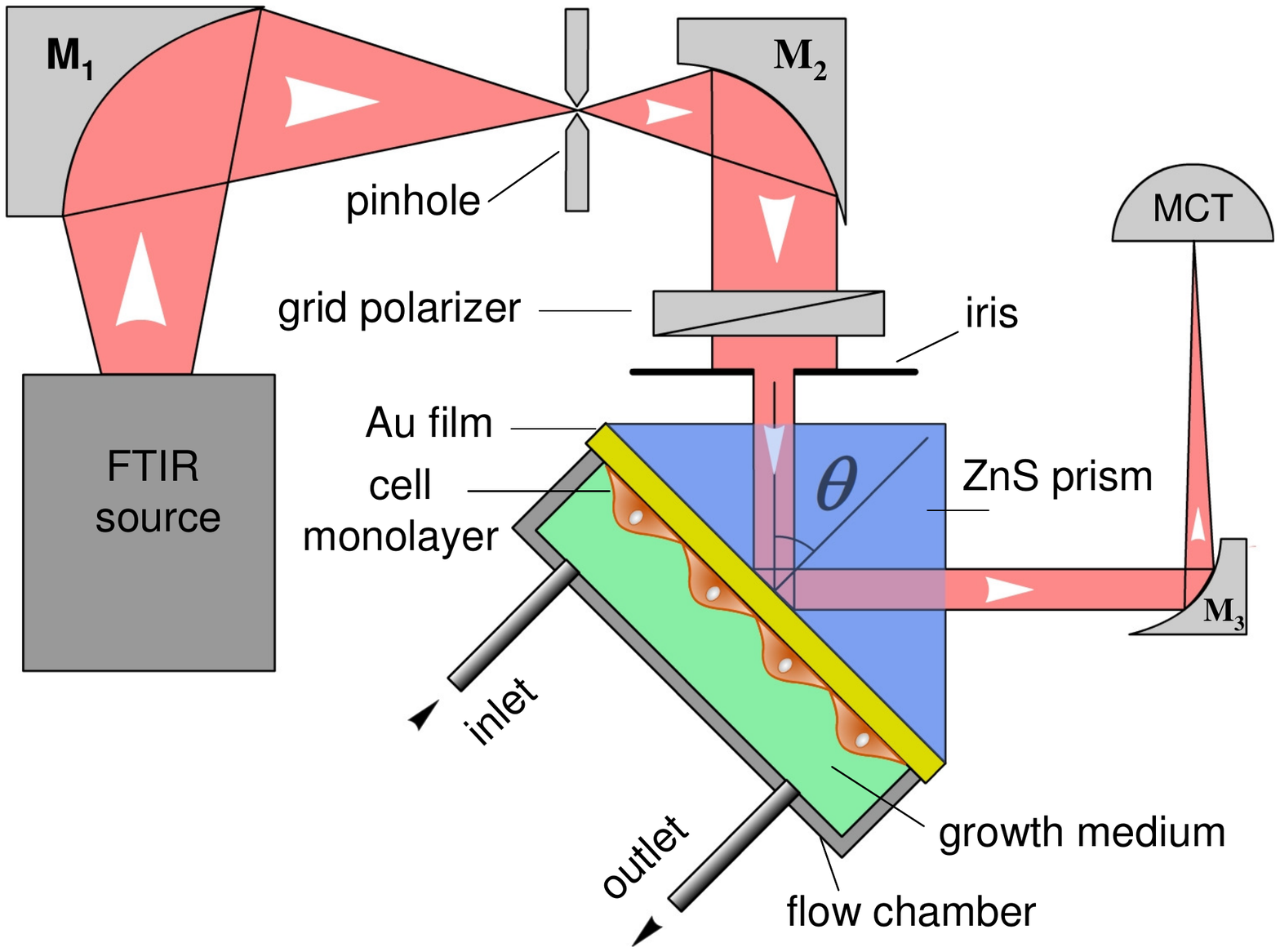}
\caption{The FTIR-SPR setup.  The infrared beam from the FTIR spectrometer is collimated by two parabolic mirrors, M1 and M2, and a pinhole. The  collimated beam is polarized by the grid polarizer, it passes through the ZnS prism, and is reflected from its base. The reflected beam is focused by the parabolic mirror, M3, on the MCT detector. The base of the prism is Au-coated and it is in contact with the flow chamber. The samples can be of two kinds: (i) a biosolution in the flow chamber, or (ii) a biological membrane or cell monolayer grown directly on the base of the Au-coated prism. In the latter case, the flow chamber is filled with growth medium.}
\label{fig:setup}
\end{figure}

Next, we will present several biological applications of the FTIR-SPR technique.  Figure \ref{fig:setup}  shows our experimental setup based on the Bruker Equinox 55 FTIR spectrometer.  The light source is a tungsten lamp  equipped with the KBr beam splitter. The infrared beam is emitted from the external port of the spectrometer and passes through a collimator consisting of a pinhole mounted at the common focal plane of two Au-coated off-axis parabolic mirrors with a focal length of 76.2 mm and 25.4 mm, correspondingly. For the 1-mm-diameter pinhole, the beam diameter is 3-4 mm and the beam divergence is $\Delta\Theta_{div}= 0.8^o$. The collimated beam passes through the grid polarizer (Specac, Ltd.) and is reflected from the right-angle  ZnS prism (ISP Optics, Inc.) mounted on a $\Theta-2\Theta$ rotating table.  Another parabolic mirror focuses the reflected beam on the liquid-nitrogen-cooled MCT (HgCdTe) detector. A temperature-stabilized flow chamber (with a volume of 0.5 ml) is in contact with the gold-coated base of the prism. The  gold film thickness is chosen according to the targeted wavelength (Fig. \ref{fig:width}, inset).

To operate this setup, we first choose the desired incident angle. Then we mount the sample and  measure the reflectivity spectrum for the $s$-polarized beam. This  spectrum is used as a background for further measurements. Then we measure the  reflectivity spectrum for the $p$-polarized beam. Normally, this is done with 4- cm$^{-1}$ resolution and 16 scan averaging. 

\subsection{Solutions}
\subsubsection{Glucose concentration in water}
\begin{figure}[ht]
\includegraphics[width=0.6\textwidth]{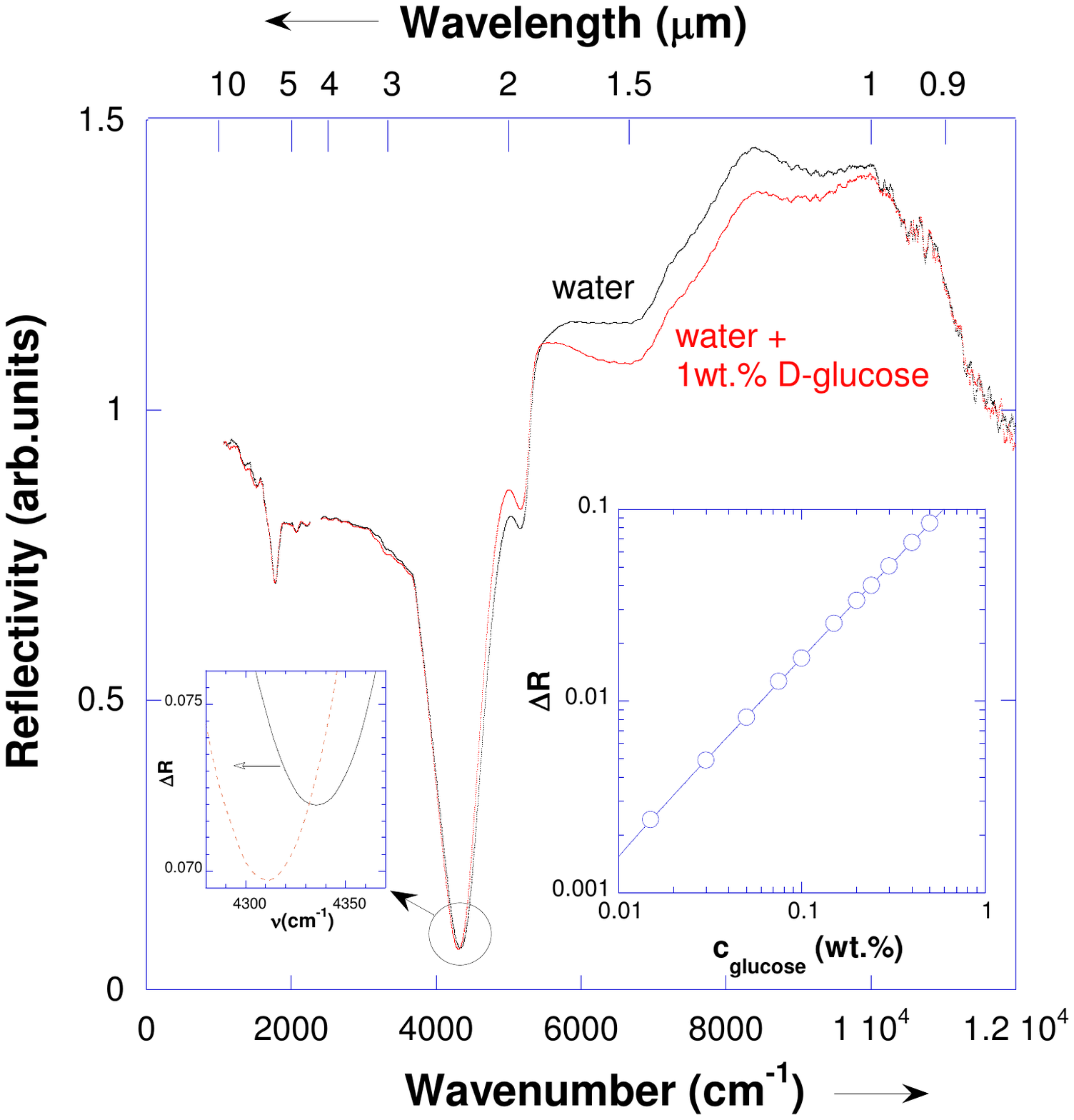}
\caption{Reflectivity  from the ZnS/Au/water  interface (black dots). The gold film thickness is 13 nm, and the incident angle is $\Theta_{ext}=21^o$. The red dots denote the corresponding results for 1 wt.$\%$ of D-glucose solution in water. The surface plasmon resonance for the Au/water interface is manifested by a pronounced dip at 4334 cm$^{-1}$. Addition of 1 wt.$\%$ of D-glucose shifts this dip to 4310 cm$^{-1}$. A secondary dip at 1780 cm$^{-1}$ is a long-wavelength surface plasmon; the features at 5145 cm$^{-1}$ and at 6900 cm$^{-1}$ indicate the water absorption peaks. The inset shows that $\Delta R$ at 4600 cm$^{-1}$ (this wavenumber corresponds to the slope of the surface plasmon resonance and the reflectivity here is extremely sensitive to slight variations in the refractive index) linearly depends on the D-glucose concentration. The slope of this linear dependence yields a bulk sensitivity, $S_{bulk}=dR/dn$=75  RIU$^{-1}$.}
\label{fig:glucose}
\end{figure}

We used the FTIR-SPR technique for precisely measuring the refractive index of solutions.  Our motivation was to monitor physiologically important glucose concentrations in water and in human plasma \cite{Lirtsman2008}. Similar studies have been performed earlier using SPR in the visible range  \cite{Lam,Aslan,Hsieh}. Figure \ref{fig:glucose} shows the reflectivity spectra of pure water and of 1$\%$ D-glucose solution measured using our setup. The spectral range  around 4000 cm$^{-1}$ corresponds to the short-wavelength SPR, and the small peak at 2000 cm$^{-1}$ corresponds to the long-wavelength SPR.  Adding glucose affects the whole reflectivity spectrum, in such a way that the SPR resonance is red--shifted. The maximal  reflectivity change occurs at 4600 cm$^{-1}$. The inset shows that the reflectivity  at $4600 cm ^{-1}$  linearly increases with increasing glucose concentration. The slope of this linear dependence yields  sensitivity, $S_{bulk}=\partial R/\partial n=75$ RIU$^{-1}$. Taking into account the actual  beam divergence of 0.8$^o$, the experimental sensitivity is consistent with the approximate expression given by Eq. \ref{S-bulk} (see also Fig. \ref{fig:sens-bulk}).

\subsubsection{Glucose uptake by erythrocytes}
Here we studied glucose uptake by erythrocyte suspension in the PBS medium \cite{buffer}.  Briefly, fresh human red blood cells (RBCs) were washed four times in PBS by centrifugation. The supernatant and buffy coat cells were discarded and RBCs were resuspended in glucose-free PBS to yield $c_{e}=5\%$ v/v. For complete glucose depletion, cells were incubated for 60 min at 22$^o$ C in PBS. Then, RBCs were centrifuged and resuspended in fresh PBS containing the required concentrations of D-glucose supplemented with L-glucose, to keep the total glucose concentration (osmolarity) equal to 20 mM. The SPR measurements were performed at 22$^o$C to slow down cellular glycolysis. 

Our measurements were performed using a ZnS prism coated with a 12-nm thick Au film. We chose the wavelength $\lambda=$4 $\mu$m at which the surface plasmon  penetration depth, $\delta _{zd}=$4.5 $\mu$m (Fig. \ref{fig:penetration-depth}), is comparable to the typical erythrocyte diameter. Although the sensitivity at this wavelength is low (Fig. \ref{fig:sens-bulk}), the use of such a long wavelength here is mandatory in order to sense the processes inside erythrocytes.  The average distance between the erythrocytes  is $\sim D_{e}c_{e}^{1/3}=$17 $\mu$m.  The surface plasmon propagation length, $L_{x}=$30 $\mu$m (Fig. \ref{fig:propagation-length}), exceeds the distance between erythrocytes. Hence, in the "coherence area" of the surface plasmon wave, ($\delta_{zd}\times L_{x}$), there are 1-2 erythrocytes.

\begin{figure}[ht]
\includegraphics[width=0.5\textwidth]{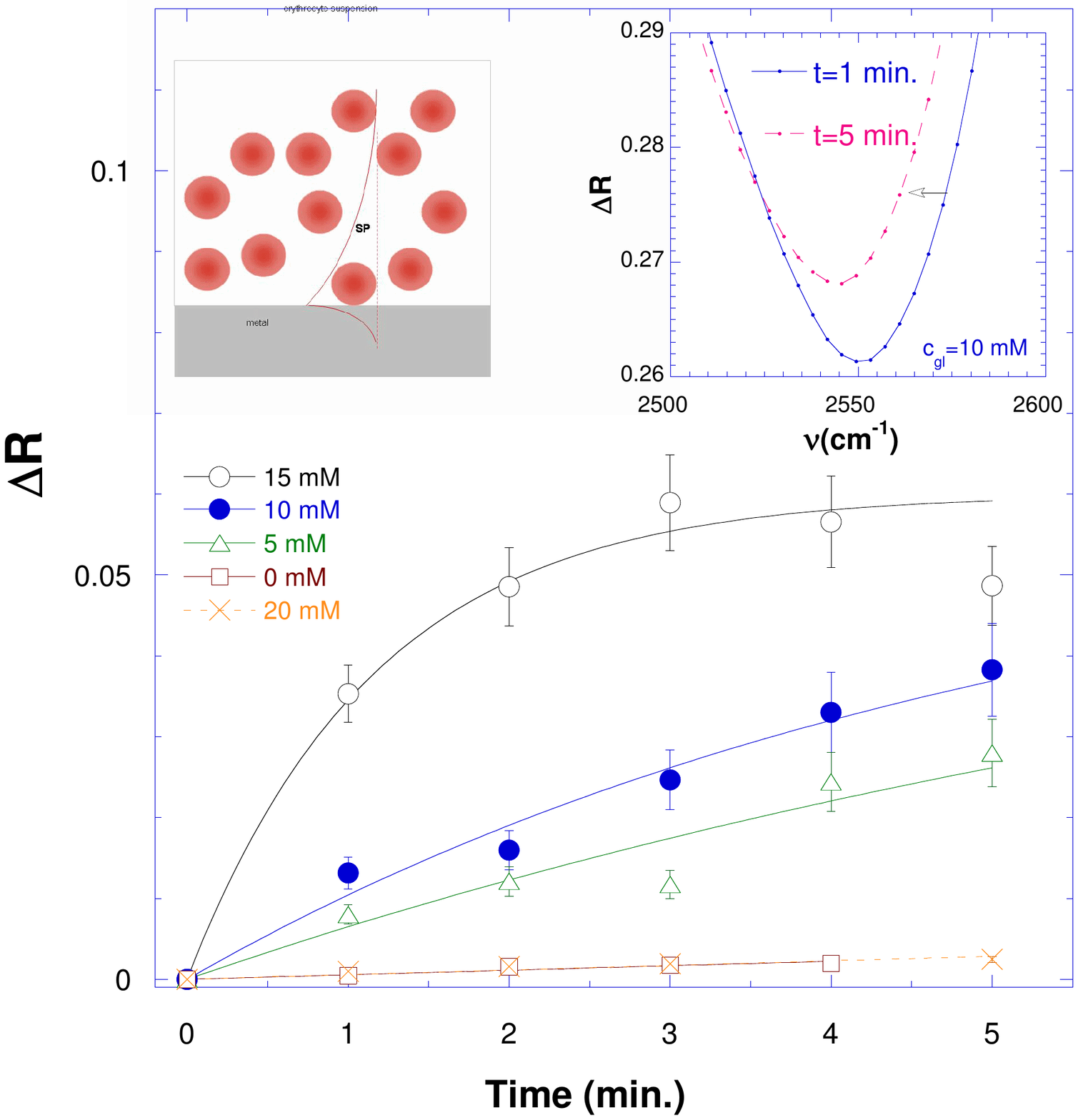}
\caption{Glucose uptake by erythrocyte suspension.  We used here a long-wavelength surface plasmon at $\lambda \approx$ 4 $\mu$m and a 12-nm-thick Au film in order to achieve a sufficiently high penetration depth, $\delta_{zd}=4.5$ $\mu$m, which is comparable to the typical erythrocyte diameter. The left-hand inset schematically shows surface plasmon penetration into erythrocytes.  The right-hand inset shows  the SPR reflectivity from the erythrocyte suspension before and after exposure to 10 mM of D-glucose. The SPR minimum shifts with time to longer wavelengths (red-shift), indicating glucose uptake by erythrocytes. The open symbols in the main panel indicate SPR reflectivity change (at a fixed wavelength) for erythrocyte suspension in the PBS buffer with different concentrations of D-glucose.  Upon addition of D-glucose, $\Delta R$ increases and achieves saturation. Continuous lines indicate exponential fit, $\Delta R\propto(1-e^{-\alpha t})$, whereas the rate $\alpha$ increases with glucose concentration. The orange crosses indicate the results of another experiment where cells were exposed to solution containing 20 mM D-glucose supplemented with 20 $\mu$M of the GLUT-1 inhibitor, cytochalasin B (CB). As expected, a negligible  reflectivity variation was observed in the presence of the inhibitor.}  
\label{fig:erythro}
\end{figure}

The erythrocytes intensively absorb D-glucose, which affects their refraction index and probably their shape. The optical reflectivity from the erythrocyte suspension  changes correspondingly.  Indeed, Fig. \ref{fig:erythro} shows that, upon addition of D-glucose,  reflectivity in the SPR regime rapidly increases with time, until saturation occurs. The rate of  this process  increases when cells are exposed to higher glucose concentrations. 

In another experiment, cells were exposed to a solution containing 20 mM D-glucose supplemented with 20 $\mu$M of the GLUT-1 inhibitor, cytochalasin B (CB). A negligible reflectivity change was observed in the presence of the inhibitor. Similarly, minor changes in  the reflectivity were detected when cells were exposed to a plain buffer solution. We interpret these results as follows: the erythrocytes absorb D-glucose (but not L-glucose), and this affects their refraction index. Although the exact mechanism by which the surface plasmon wave interacts with the erythrocytes is not yet clear, we believe that the observed SPR variation is associated with D-glucose uptake by erythrocytes because (i) the rate of  SPR change is determined by the glucose concentration and (ii) the SPR change is inhibited by cytochalasin B, as expected.  Therefore, we can conclude that FTIR-SPR measurements specifically sense a glucose-dependent transport process. This conjecture is corroborated by the dielectric spectroscopy measurements on similar suspensions \cite{Livshitz}. 

\subsection{Cell culture studied by the SPR - the physics involved} 
 We succeeded in growing cell cultures directly on the Au-coated prism \cite{Ziblat,Yashunsky}. These include human Melanoma (MEL 1106), MDCK, and HeLa cells that form monolayers with a typical confluence of 60-80$\%$.
In a typical experiment, the prism with a cell culture was removed from the incubator, attached to a flow chamber, mounted in the FTIR-SPR setup, and then exposed to plain buffer medium (MEM) at 37$^o$C. The temperature in the flow chamber was controlled within 0.1$^o$C while the buffer solution was constantly pushed through the chamber at a flow rate of 5 $\mu$l/min  using a motorized bee syringe pump. After 5-7 minutes of equilibration, we measured the reflectivity in the SPR regime. The measurements  lasted up to 6 hours, after which, the cells usually die, most probably from the lack of CO$_{2}$.

\subsubsection{Surface plasmon propagation in the cell culture}
Understanding surface plasmon propagation in the cell monolayer necessitates further discussion. Indeed, the cells have irregular shapes with a lateral size of $\sim10\times 20$ $\mu$m$^2$ and a  height of 1 to 6 $\mu$m. Since the cells contain up to 25$\%$ organic molecules, their average refractive index exceeds that of water. This results in an angular shift of the SPR resonance (Eq. \ref{Theta-SP}) for the ZnS/Au/cell interface, compared with the ZnS/Au/water interface. However, even at high cell confluence, some parts of the Au-coated prism are still uncovered by cells. These Au patches may exhibit an unshifted SPR, provided their size  exceeds the SP propagation length,  $l_{patch}>L_{x}$.

\begin{figure}[ht]
\includegraphics[width=0.6\textwidth]{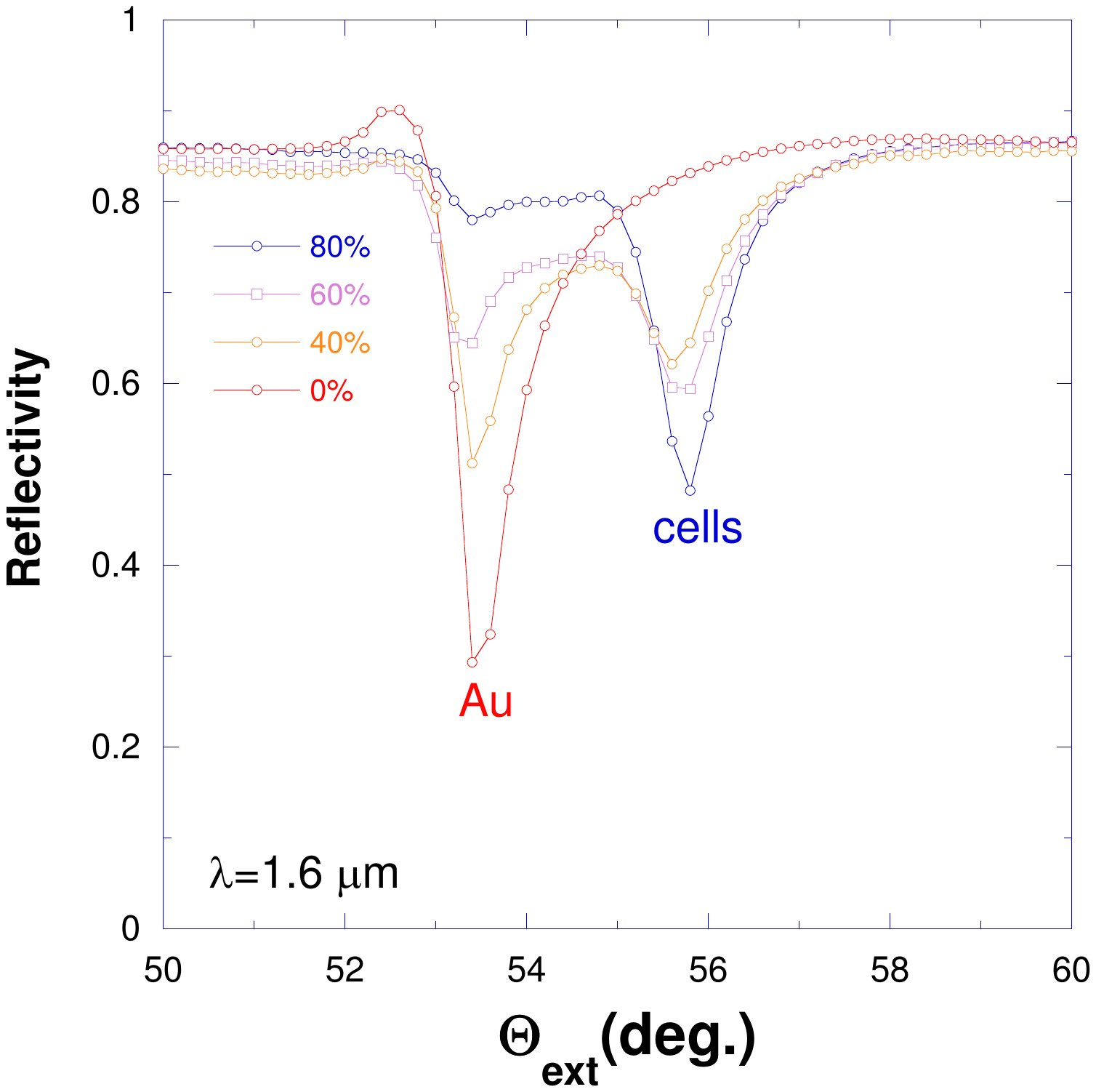}
\caption{Angular dependence of the surface plasmon resonance from the HeLa cell monolayers with  different cell coverage (confluence). We used here an SF-11 right-angle glass prism coated with a 35-nm-thick Au film. In the absence of cells, there is a single SPR minimum at $\Theta_{ext}=53.4^o$ that corresponds to the reflectivity from the ZnS/Au/water interface. In the presence of cells, an additional SPR dip appears at $55.8^o$  that corresponds to the reflectivity from the ZnS/Au/cells interface. The angular shift between these two minima yields the refractive index difference, $n_{cell}-n_{water}=0.03$. At high cell confluence (80$\%$), the SPR signal from the cells is dominant.}
\label{fig:HeLa}
\end{figure}

Figure \ref{fig:HeLa} summarizes our studies of the HeLa cell monolayers grown on an SF-11 gold-coated glass prism. Here  we used a relatively short wavelength $\lambda=$1.6 $\mu$m. The  corresponding penetration depth is also small, $\delta _{zd}=$0.7 $\mu$m (Fig. \ref{fig:penetration-depth}) but sufficient to probe the cell's interior. The propagation length, $L_{x}=$ 46 $\mu$m (Fig. \ref{fig:propagation-length}), exceeds the average cell size. Figure \ref{fig:HeLa} shows angular-dependent reflectivity for different cell coverages (confluences). In the absence of cells, there is a single dip at $\Theta_{sp}=$ 53.4$^o$ that corresponds to the surface plasmon resonance at the ZnS/Au/water interface. In the presence of HeLa cells,  an additional dip appears at  $\Theta_{sp}=$ 55.8$^o$ that corresponds  to the ZnS/Au/HeLa cell interface. The SPR angular shift corresponds to  $n_{cell}-n_{water} =$ 0.03. For high cell confluence, e.g., 80$\%$, the SPR from uncovered gold patches is barely seen, as expected.

\subsubsection {Guided modes}
\begin{figure}[ht]
\includegraphics[width=0.8\textwidth]{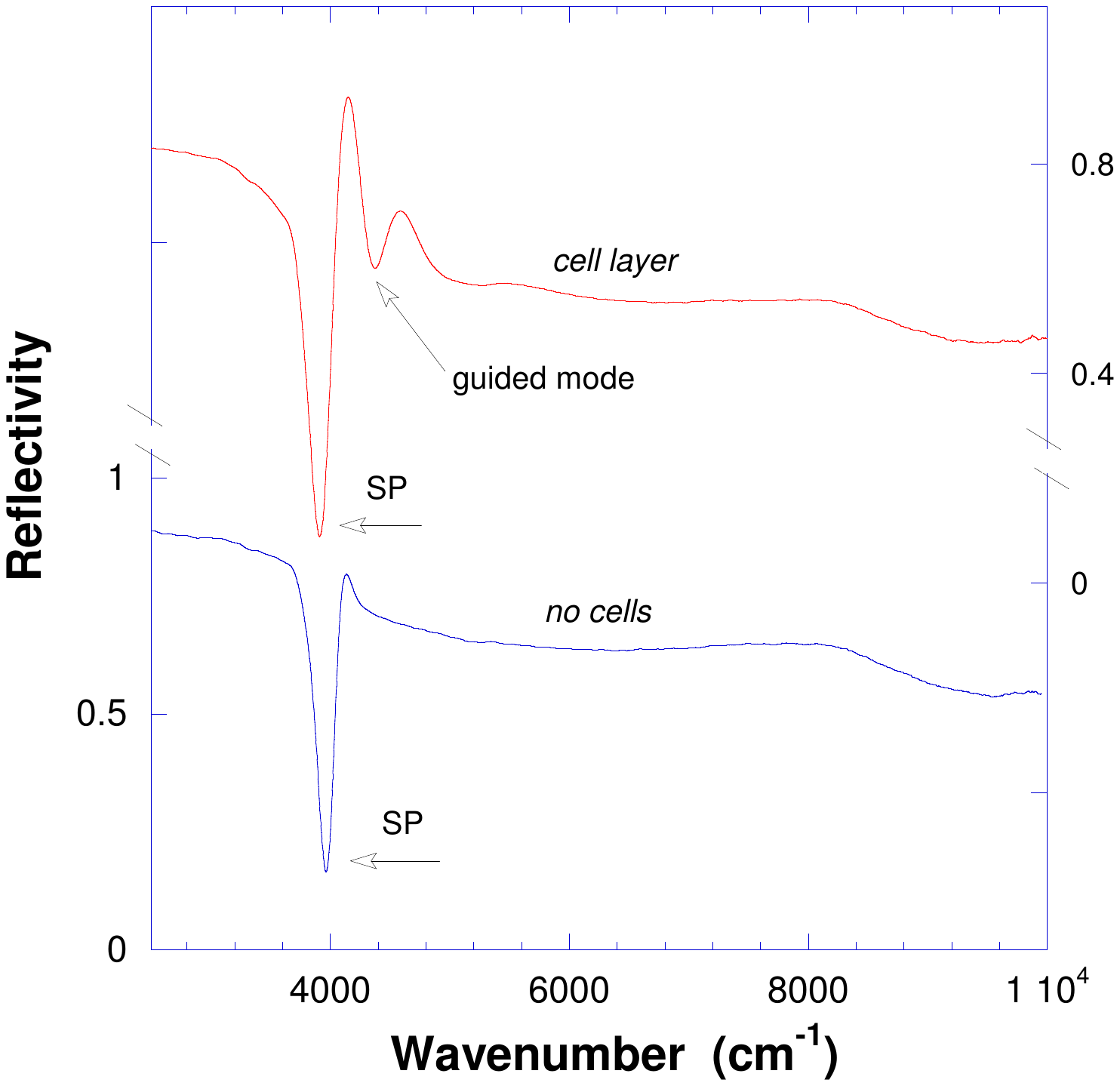}
\caption{Reflectivity in the SPR regime for  a Melanoma 1106-cell culture in MEM growth solution (upper curve, $\Theta=$ 19.5$^o$), and  growth solution without cells (lower curve, $\Theta=$ 18.6$^o$).  The upper curve is displaced upwards for clarity. Note the main SPR dip at 3920 cm$^{-1}$ in both curves and a satellite dip at 4385 cm$^{-1}$ in the upper curve.  We associate this dip with the $TE_{0}$ waveguide mode propagating in the cell monolayer.} 
\label{fig:waveguide}
\end{figure}

Since the refractive index of cells exceeds that of the surrounding aqueous medium ($n_{cell}>n_{water}$), and the cell monolayer thickness is comparable to the mid-infrared wavelength ($d_{cell} \sim \lambda$), the cell monolayer on gold could behave as a metal-clad optical waveguide \cite{Yeh}.  An indication of such behavior is seen in Fig. \ref{fig:waveguide}, which shows an SPR spectra  (i) for a high-confluence human MEL 1106 cell monolayer grown on an Au-coated ZnS prism, and (ii) for the similar Au-coated prism without cells. In both cases, there is a strong surface plasmon resonance at 3920 cm$^{-1}$ ($\lambda=2.55$ $\mu$m). For the prism with cell culture,  a short-wavelength satellite appears at  4385 cm$^{-1}$ ($\lambda=2.28$ $\mu$m). We associate this feature with the waveguide mode propagating in the cell monolayer (Fig. \ref{fig:cells}d). 

Indeed, the  cut-off condition for the waveguide modes in the planar metal-clad dielectric waveguide is \cite{Knoll,Yeh}
\begin{equation}
k_{0}d_{cell}\sqrt{n_{cell}^2-n_{water}^2}=\left(m-\frac{1}{2}\right)\pi,
\label{cutoff}
\end{equation}
where $m=1,2...$. We substitute here $n_{water}=1.25$, $n_{cell}-n_{water}=0.03$, $d_{cell}=2$ $\mu$m; for the lowest $TM_{1}$ mode we found  $\lambda_{0}$= 2.3 $\mu$m ($\nu=4348$ cm$^{-1}$). This estimate corresponds fairly well to our observation shown in Fig. \ref{fig:waveguide}.  Note that the surface plasmon propagation length at this wavelength is very long, $L_{x}=73$ $\mu$m (Fig. \ref{fig:propagation-length}), as expected. 

\subsection{Cell cultures studied by the SPR technique - the biology involved} 
\subsubsection{Cholesterol in the cell membrane} 
We studied cholesterol penetration into the plasma membrane of the HeLa cells \cite{Ziblat} using our FTIR-SPR technique (similar studies using SPR in the visible range were performed in Ref. \cite{chol2}). It is well-known that cholesterol enters mostly into the plasma membrane rather than into cytoplasm. Therefore, to achieve high sensitivity to  membrane-related events, we chose the near-infrared wavelength range, $\lambda_{sp}=1-1.08$ $\mu$m, characterized by a relatively low SPR penetration depth. 

The HeLa cells were grown on an SF-11 glass prism coated with 35-nm-thick gold film. The prism with cells  was mounted into a flow chamber and was equilibrated in growth medium  for 5-7 minutes  at 37$^o$ C. Then we added 10 mM of m$\beta$CD-chol (methyl-beta-cyclodextrin loaded with cholesterol), which is known to enrich the cells by cholesterol. 

\begin{figure}[ht]
\includegraphics[width=0.6\textwidth]{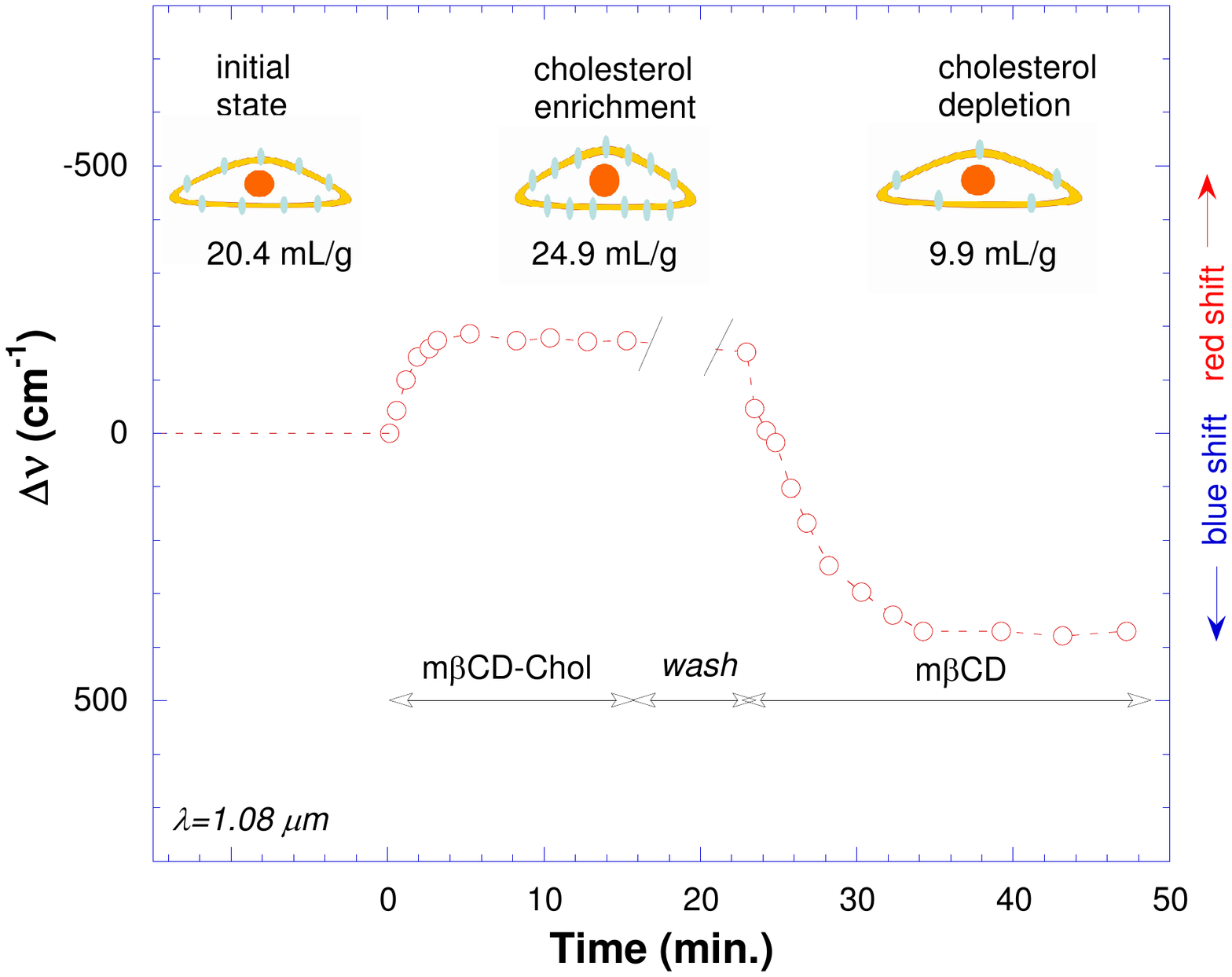}
\caption{Surface plasmon resonance shift upon cholesterol enrichment/depletion of HeLa cell membranes.  The drawings schematically show cholesterol molecules in the cell membrane (gray ovals); the numbers below the drawings indicate cholesterol concentration in the membrane, as was determined biochemically on analogous cell cultures.  In order to be sensitive to membrane-related processes, the SPR measurements were performed at a short wavelength, $\lambda=1.08$ $\mu$m, to ensure short penetration depth, $\delta_{zd}=0.36$ $\mu$m (see Fig. \ref{fig:penetration-depth}).  At $t=0$  we injected 10 $\mu$M of m$\beta$CD-chol solution. The SPR becomes red-shifted, as expected, since the accumulation of cholesterol  molecules with a high refractive index in the cell membrane increases the average refractive index. At $t=$15 min, the cells were washed with plain medium for 5 min and then  3 $\mu$M  m$\beta$CD solution was added.  The SPR became blue-shifted since this drug depletes cell membranes from cholesterol in such a way that the average refractive index in the SPR-sensed region decreases.} 
\label{fig:cholesterol}
\end{figure}

We measured FTIR-SPR spectra before and after adding the drugs.
Figure \ref{fig:cholesterol} shows that after exposure to m$\beta$CD-chol,  the SPR is red-shifted. This corresponds to a refractive index increase because cholesterol has a higher refractive index than water. In contrast, when we add to a growth solution a similar chemical, m$\beta$CD,  which is not loaded with cholesterol, the SPR is blue-shifted, which corresponds to a refractive index decrease. This is expected since  m$\beta$CD  depletes plasma membranes from cholesterol. Consistent with this explanation, the SPR reflectivity in the cholesterol-depleted state is lower than  the initial SPR reflectivity.

\subsubsection{Ferrotransferrin uptake}
\begin{figure}[ht]
\includegraphics[width=0.8\textwidth]{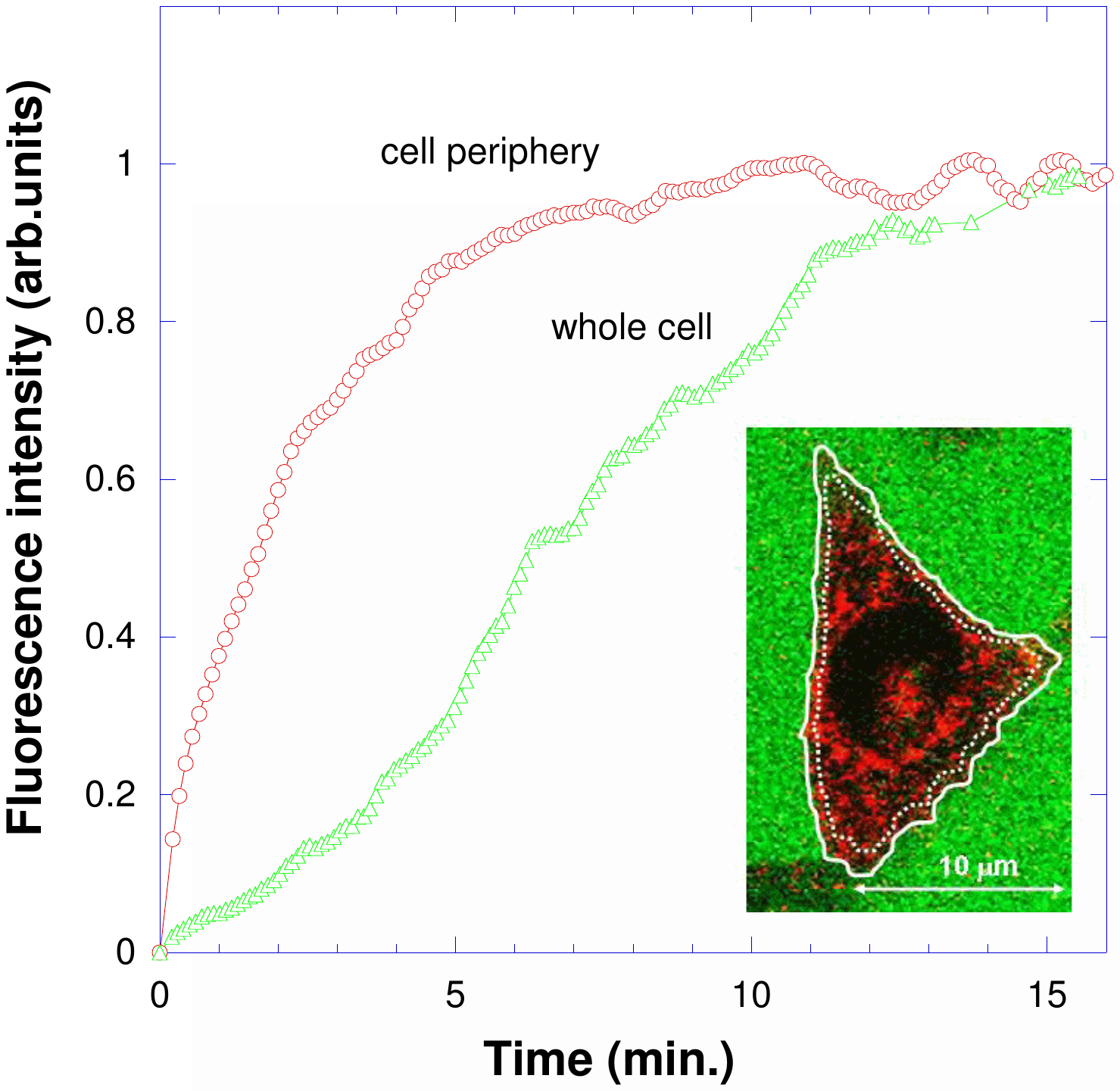}
\caption{Cell fluorescence resulting from the Rhodamine-tagged  ferrotransferrin (Ft)penetration into cells. The red circles represent the peripheral region of the cell, whereas the green triangles denote the fluorescence from the whole cell. Note the fast fluorescence  kinetics in the peripheral regions of the cell, as compared with the slow kinetics in the  whole cell. The inset shows the confocal microscope image of a single cell under partial penetration of the ferrotransferrin.} 
\label{fig:fluorescence}
\end{figure}

We applied our FTIR-SPR technique to study transferrin-induced clathrin-mediated endocytic processes that introduce Fe ions into the cell.  Figure \ref{fig:fluorescence} shows a fluorescence image of the Melanoma 1106 cell obtained by confocal microscopy \cite{confocal}. The inset shows the cell interior in black; the red areas indicate Fe penetration, and the green areas indicate extracellular space. It is clearly seen that the fluorescence intensity increases and then achieves saturation, whereas its kinetics in the peripheral part of the cell is faster than that in the whole cell. 

\begin{figure}[ht]
\includegraphics[width=0.6\textwidth]{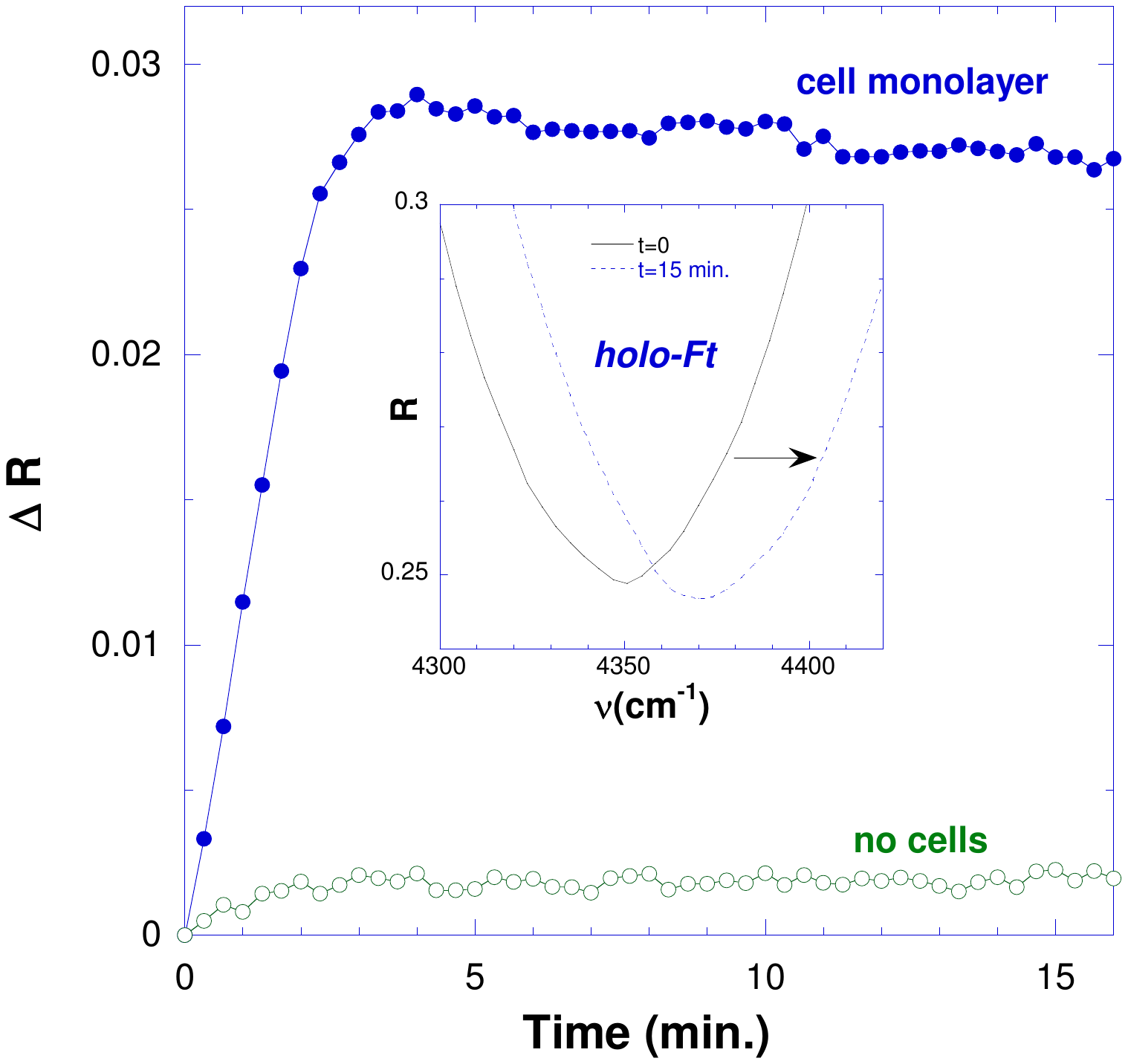}
\caption{Reflectivity variation in the SPR regime following the introduction of holo-Ferrotransferrin. The open circles indicate the results for the  ZnS/Au/growth solution interface when there is no cell culture.  Here, introduction of the holo-Ft does not result in significant changes in reflectivity, indicating negligible holo-Ft adsorption on the Au surface. The filled circles indicate a holo-Ft-induced reflectivity change from the ZnS/Au/Mel 1006/water interface. There is pronounced SPR shift that closely follows the kinetics of fluorescence from the peripheral parts of the cells. The inset shows that the SPR in this case is blue-shifted.} 
\label{fig:holo}
\end{figure}

We studied the same process using a long-wavelength surface plasmon at $\lambda=$2.54 $\mu$m. Its penetration depth, $\delta_{zd}=1.2$ $\mu$m (Fig. \ref{fig:penetration-depth}), is long enough to penetrate the cells, although it senses mostly the cell periphery.  Figure \ref{fig:holo} shows  SPR reflectivity variation upon introduction of holo-Ft into the solution. The kinetics of the SPR reflectivity closely follows  the  fluorescence kinetics in the cell periphery (Fig. \ref{fig:fluorescence}), as expected. This example demonstrates that the SPR results are consistent with the confocal microscopy observations using fluorescent tags. The obvious advantage of the SPR technique is that it is label-free. 

Interestingly,  the SPR here is blue-shifted, indicating that the average refractive index in the measured volume decreases (Fig. \ref{fig:holo}, inset). This rules out the possibility that the SPR shift results from accumulation of organic molecules in the cell (this would increase the refractive index). The blue shift indicates that the cells become "diluted", as if the growth solution penetrates into cells. This is consistent with the biological picture of  endocytosis  that includes transferrin-induced vesicular transport.

\section{Discussion and Conclusions}
\begin{itemize}
    \item In the context of biosensing, the lack of convenient laser sources and of enhanced water absorption were considered as major impediments for  extending the surface plasmon resonance  technique into the mid-IR range. We demonstrated here that there are several spectral windows, whereby clever choice of  wavelength, beam collimation, angle of incidence, and metal film thickness, the SPR technique can achieve sensitivity that is comparable and even higher than that in the visible range. 
    \item Surface plasmon resonance in mid-IR  can be used to study cells and cell cultures. This technique is well matched  for the sensing of cells since the surface plasmon  penetration depth in the infrared range is just on the order of the cell size. This is in contrast to the surface plasmon  in the visible range, whose penetration depth is too short to probe the processes inside the cell cytoplasm.  However, to develop the FTIR-SPR technique into a useful tool for studying cells, one must theoretically analyze surface plasmon propagation and scattering in a cell monolayer. 
    \item The waveguide modes in cell culture may develop into a useful tool to study cell-cell attachment  and cell adhesion to substrates.
    \item The most challenging task in the field of FTIR-SPR  is to convert it into a spectroscopic tool that allows measuring the FTIR absorption/dispersion spectra in the regime of the surface plasmon resonance. The sensitivity of this technique should surpass that of FTIR-ATR  \cite{Chittur} and  allow detection of biomolecules according to their "fingerprints".
\end{itemize}

\section{Acknowledgments} 
We are grateful to  S. Shimron for preparing the cell cultures and  for significantly contributing to the SPR experiments associated with them, to L. Livshitz for his help in erythrocyte studies, as well as to N. Melamed-Book and A. M. Weiss for confocal microscope images and processing. This study was supported in part by the Israel Science Foundation (Grant No. 1337/05),  by Johnson $\&$ Johnson, Inc., and by a Bikura grant.
\pagebreak
 {}
\pagebreak

\begin{thebibliography}{}
\bibitem{buffer} The phosphate-buffered saline buffer (PBS) has a  pH of 7.4 and contains 137 nM NaCl, 8.9 mM Na$_{2}$HPO$_{4}$, 2.7 mM KCl, 1.5 mM KH$_{2}$PO$_{4}$. We used (from healthy donors) fresh human erythrocytes that were washed in PBS and centrifugated three times for 5 min at 300x$g$. The washed erythrocytes were suspended in the PBS and stored in ice.
\bibitem{confocal} The cells were imaged with an Olympus FV-1000 confocal microscope, equipped with an on-scope incubator (Life Image Services). The holo-Ft was tagged with Rhodamine Red.
\end{thebibliography}
\end{document}